\documentclass[11pt,aps,amsmath,amssymb,preprint,nofootinbib,eqsecnum]{revtex4}
\usepackage[active]{srcltx}
\usepackage[utf8]{inputenc}
\usepackage{latexsym}
\usepackage{amsmath}
\usepackage{graphicx}

\begin{document}

\title{Effective Theories for 2+1 Dimensional Non-Abelian Topological Spin Liquids}

\author{Carlos A. Hernaski}
\email{carlos.hernaski@gmail.com}
\affiliation{Departamento de F\'isica, Universidade Estadual de Londrina, 
Caixa Postal 10011, 86057-970, Londrina, PR, Brasil}

\author{Pedro R. S. Gomes}
\email{pedrogomes@uel.br}
\affiliation{Departamento de F\'isica, Universidade Estadual de Londrina, 
Caixa Postal 10011, 86057-970, Londrina, PR, Brasil}

%%%%%%%%%%%%%%%%%%%%%%%%%%%%%%%%%%%%%%%
\begin{abstract}

In this work we propose an effective low-energy theory for a large class of 2+1 dimensional non-Abelian topological spin liquids whose edge states are  conformal degrees of freedom with central charges corresponding to the coset structure $su(2)_k\oplus su(2)_{k'}/su(2)_{k+k'}$. For particular values of $k'$ it furnishes the series for unitary minimal and superconformal models. These gapped phases were recently suggested to be obtained from an array of one-dimensional coupled quantum wires. In doing so we provide an explicit relationship between two distinct approaches: quantum wires and Chern-Simons bulk theory. 
We firstly make a direct connection between the
interacting quantum wires and the corresponding conformal field theory at the edges, which turns out to be given in terms of chiral gauged WZW models. Relying on the bulk-edge correspondence we are able to construct the underlying non-Abelian Chern-Simons effective field theory.

\end{abstract}
\maketitle

%%%%%%%%%%%%%%%%%%%%%%%%%%%%%%%%%%%%%%%%%%%%%%%%%%%%

\section{Introduction}

\subsection{Motivation}

Quantum matter under strong coupling produces remarkable physical phenomena. 
An imprint of such a system is the fractionalization of quantum numbers like electric charge and statistics, 
making evident the nonperturbative nature of the problem. We need to go beyond 
traditional perturbative methods in order to fully understand their properties.  
In addition, they do not fit in the general symmetry-breaking scheme for classifying phases, since  
there is no local order parameter able to distinguish phases. Rather, they can be classified according to topological 
properties being generically designated as topological phases of matter. 

It is a great challenge to classify and organize in a common framework the description of the internal orders of topological phases. 
A major achievement in this direction was the perception of ubiquitous features shared for all these systems like the existence of gapless edge states, bulk anyonic quasi-particle excitations, topological degeneracy, etc \cite{Wen}. Within the class of $2+1$ dimensions gapped topological phases, the only gapless excitations reside at the boundary of the manifold. Then the low energy limit of the system should be described by a two dimensional conformal field theory (CFT). The observation by Witten \cite{Witten} that current algebra in two dimensions is equivalent to a three dimensional Chern-Simons theory suggested a possibility to classify all rational CFT by a bulk Chern-Simons theory. In \cite{Seiberg}, this connection is made explicit by showing the equivalence of the Hilbert spaces of general Chern-Simons theories with suitable gauge groups with those of the Wess-Zumino-Witten (WZW) models and its gauged version. This characterization scheme for the CFT also provided a method for the description of the effective low-energy theory for the gapped topological phases in $2+1$ dimensions in terms of the Chern-Simons topological field theory \cite{Wen,Wen2,Wen3,Fradkin,Nayak,Carrega}.

Progress in the characterization of topological phases has also been done in terms of a distinct framework denominated quantum wires. 
In this approach the 2+1 dimensional phase is thought as a limit of a highly anisotropic system constituted of 
1+1 dimensional wires arranged in a periodic way. This system describes a CFT in $1+1$ dimensions characterized by a certain central charge proportional to the number of wires. By switching on electron-electron fine-tuned interactions of finite range, one is able to render a mass gap to the bulk electrons, consequently reducing the total central charge of the system. The only gapless excitations reside on the two extreme bundles of quantum wires.
In addition to offer in a certain sense a more microscopic view of the topological phase, a great advantage of this construction 
is to have at our disposal all the machinery of bosonization of two dimensional quantum field theories. 
Fractional quantum Hall phases were successfully constructed in \cite{Kane,Teo,Jelena} and further generalization to other phases were
presented  in \cite{Neupert,Huang}.

The main purpose of the present work is to make an explicit connection between the 
two mentioned approaches by proposing an effective bulk theory for the non-Abelian spin liquids constructed within the quantum wires scheme.

%%%%%%%%%%%%%%%%%%%%%%%%%%%%%%%%%%%%%%%%%%%%%%%%%%%%%%%
\subsection{Main Results}

The work of Ref. \cite{Huang} gives a low-energy description of classes of non-Abelian topological spin liquids in terms of an array of quantum wires coupled through current-current interactions. Based on the bulk-edge correspondence \cite{Wen2, Wen3} one could expect that the effective theory for this topological phase should be expressed in terms of a non-Abelian Chern-Simons theory, which is then seen as a bulk theory for the quantum wires system. In our analysis we have suggested how this connection can be made explicit. Namely, we started from the quantum wires and we have shown that its strong coupling limit is given in terms of a gauged WZW model at the edge, which in turn corresponds to a bulk non-Abelian Chern-Simons theory. In this process we argued in favor of the plausibility in trading the strong coupling limit of the interactions among neighboring wires by direct constraints on the currents. In the work of Ref. \cite{Cabra} it is shown that such a set of constrained free fermions in $1+1$ dimensions is a realization of a gauged WZW model, which implements the Sugawara construction (see, for example, Ref. \cite{Francesco}) of the conformal field theories for a coset of simple groups known as GKO construction \cite{GKO}. This relation paves the way for our identification of the strongly coupled quantum wires system with the gauged WZW model. Notwithstanding, this is not an arbitrary representation of the edge conformal field theory, but it follows directly from the physical setup of quantum wires. In Ref. \cite{Seiberg} one establishes a general method to implement the correspondence between a $1+1$ dimensional conformal field theory and a $2+1$ dimensional Chern-Simons topological field theory, first emphasized in \cite{Witten}. This method is used in Ref. \cite{Fradkin} to construct an explicit realization of the non-Abelian topological phase for the Pfaffian fractional quantum Hall state \cite{Moore-Read} in terms of a bulk Chern-Simons theory with suitable gauge groups glued by appropriated boundary conditions. In the present work we implement this construction to describe the non-Abelian spin liquid topological phases with the central charge associated with the coset structure of Lie-algebras
\begin{equation}
\left(\frac{su(2)_k\oplus su(2)_{k'}}{su(2)_{k+k'}}\right)_R~~~\text{and}~~~\left(\frac{su(2)_k\oplus su(2)_{k'}}{su(2)_{k+k'}}\right)_L,
\label{1.1}
\end{equation}
in terms of a bulk Chern-Simons theory. With this effective theory one can in principle explore important effects of the topological phase, like the braiding statistics of the non-Abelian anyons and the topological degeneracy of the ground state. We believe that our whole construction can be extended to other classes of topological phases with important phenomenological consequences.

This work is organized as it follows. In Sec. \ref{Sec1} we review certain aspects of the coupled quantum wires construction, which are useful in the remaining of the paper. Sec. \ref{Sec2} is devoted to the identification of the conformal field theory of the edge in terms of constrained fermions. In Sec. \ref{CGWZ} we discuss the equivalence between the constrained fermions theory and the gauged WZW theory, which is suitable for the connection with the bulk theory. In Sec. \ref{Sec3} we construct the bulk effective theory and establish the connection with the edge theory by means of the bulk-edge correspondence. The paper is closed with a summary and additional remarks in Sec. \ref{Sec6}. 
We have included two appendices containing a couple of auxiliary calculations.

%%%%%%%%%%%%%%%%%%%%%%%%%%%%%%%%%%%%%%%%%%%%%%%%%%%
\section{Quantum Wires Construction}\label{Sec1}

In order to make the work reasonably self-contained we start by reviewing aspects of the quantum wires approach \cite{Huang}, which are relevant to our purposes. The key idea is to describe a topological phase in 
2+1 dimensions from a set of coupled theories in 1+1 dimensions. 

\begin{figure}[!h]
\centering
\includegraphics[scale=0.7]{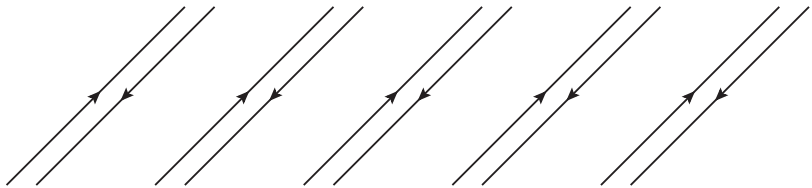}
\caption{Physical setup of wires system. The arrows represent electrons propagating in both right and left ways.}
\label{WiresPicture}
\end{figure}

The starting point is a set of spinfull electrons propagating along one dimension in both right and left ways, as represented in 
Fig. \ref{WiresPicture}. 
The corresponding Lagrangian is given in terms of free complex spinors $\psi_{L,\sigma,I}$ and $\psi_{R,\sigma,I}$,  
\begin{equation}
\mathcal{L}_0= \sum_{I=1}^N \sum_{\sigma=1}^2 \left[\mathrm{i}\psi_{R,\sigma,I}^{\ast}(\partial_t+\partial_x)\psi_{R,\sigma,I} 
+\mathrm{i}\psi_{L,\sigma,I}^{\ast}(\partial_t-\partial_x)\psi_{L,\sigma,I}\right].
\label{2.1}
\end{equation}
The index $I$ specify the wire on which the electrons are propagating.
From the physical point of view, this index can be interpreted as a ``discretized" dimension. 
The spin is specified by the index $\sigma=1,2$. The interactions will allow for electrons to pass from one wire to another.

The Lagrangian (\ref{2.1}) is a free conformal field theory with total central charge 
\begin{equation}
c= 2N,
\label{2.2}
\end{equation}
which essentially counts the number of gapless degrees of freedom. 
The strategy is to introduce interactions between neighboring wires in order to reduce part of the central charge by giving 
a gap for the wires of the bulk while leaving gapless 
the wires at the ends. 
Thus the interaction effectively turns the system into a two-dimensional one, since it enables the electrons to pass from one wire (or a set of wires) 
to another. This is the physical mechanism responsible for the production of nontrivial topological phases from quantum wires.

The symmetries preserved by the interactions determine the symmetries of the resulting topological phase.  
The Lagrangian (\ref{2.1}) is time-reversal invariant, and it is also invariant under $U_R(2N)\times U_L(2N)$, 
\begin{equation}
\psi_{R,\sigma,I} \rightarrow \psi'_{R,\sigma,I}= (U_R)_{\sigma I,\sigma' I'}\, \psi_{R,\sigma',I'}~~~\text{and}~~~
\psi_{L,\sigma,I} \rightarrow \psi'_{L,\sigma,I}= (U_L)_{\sigma I, \sigma'I'}\, \psi_{L,\sigma',I'}.
\label{2.3}
\end{equation}

\begin{figure}[!h]
\centering
\includegraphics[scale=0.7]{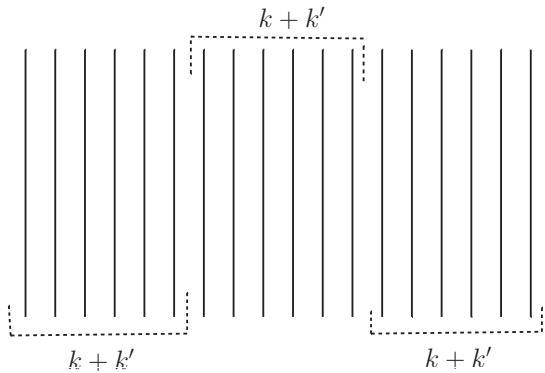}
\caption{System of wires decomposed into bundles. The picture shows three bundles constituted of $k+k'=6$ wires.}
\label{Wiresdec}
\end{figure}

In order to take advantage of the coset construction, 
we decompose the system of wires into $\mathcal{N}$ bundles 
of $k+k'$ wires, such that the total number of wires is $N=\mathcal{N}(k+k')$, as illustrated in  Fig. \ref{Wiresdec}. 
In this case the wire location of each fermion will be specified according to 
$\psi_{R/L,\sigma,I}\rightarrow \psi_{R/L,\sigma,i}^m,\psi_{R/L,\sigma,i'}^m$, 
with $m=1,...,\mathcal{N}$ running over the bundles and $i=1,...,k$ and $i'=1,...,k'$ running over wires inside a specific bundle.
The Lagrangian (\ref{2.1}) is then written as 
\begin{eqnarray}
\mathcal{L}_0&=& \sum_{m=1}^{\mathcal{N}} \left(\sum_{\sigma=1}^2\sum_{i=1}^k \left[\mathrm{i}\psi_{R,\sigma,i}^{m\ast}(\partial_t+\partial_x)
\psi_{R,\sigma,i}^m 
+\mathrm{i}\psi_{L,\sigma,i}^{m\ast}(\partial_t-\partial_x)\psi_{L,\sigma,i}^m\right]\right.\nonumber\\
&+&\left. \sum_{\sigma=1}^2\sum_{i'=1}^{k'}\left[\mathrm{i}\psi_{R,\sigma,i'}^{m\ast}(\partial_t+\partial_x)\psi_{R,\sigma,i'}^m 
+\mathrm{i}\psi_{L,\sigma,i'}^{m\ast}(\partial_t-\partial_x)\psi_{L,\sigma,i'}^m\right]\right).
\label{2.1a}
\end{eqnarray}
This form enables us to consider the subgroup 
\begin{equation}
\underbrace{ U(2k) \times U(2k')}_{\text{first bundle}} \times \cdots \times \underbrace{U(2k) \times U(2k')}_{\text{last bundle}}
\label{2.4}
\end{equation}
of $U(2N)$ for both right and left groups. 
Each one of the small groups $U(2k)$ has an associated Lie algebra $u(2k)_1$, which in turn can be decomposed as
\begin{equation}
u(2k)_1\supset u(1)\oplus su(2)_k\oplus su(k)_2
\label{2.5}
\end{equation}
and a corresponding expression for $k\rightarrow k'$. 

According to the Sugawara construction (see, for example, Ref. \cite{Francesco}), the 
corresponding energy-momentum tensor is also decomposed into the sum of the parts
\begin{equation}
T_{R/L}[u(2k)]=T_{R/L}[u(1)]+T_{R/L}[su(2)_k]+T_{R/L}[su(k)_2].
\label{2.6}
\end{equation}
A quick check for this can be done by considering the central charge of both sides. 
For an $u(2k)$ theory the central charge is $2k$. 
For the right-hand side, the central charge is the sum
\begin{equation}
1+ \frac{3k}{k+2}+\frac{2(k^2-1)}{k+2}.
\label{2.7}
\end{equation}
The energy-momentum tensor of the $u(2k)$ group is that of a free theory
\begin{equation}
T_{R/L}[u(2k)]=\frac{\mathrm{i}}{2\pi}\sum_{\sigma=1}^2 \sum_{i=1}^2 \psi_{R/L,\sigma,i}^{\ast} (\partial_t\mp \partial_x)\psi_{R/L,\sigma,i}.
\label{2.7a}
\end{equation}
The energy-momentum tensors in the right-hand-side of (\ref{2.6})
are given in terms of the underlying currents
\begin{equation}
T_{R/L}[u(1)]= \frac{1}{4k} J_{R/L}J_{R/L},
\label{2.8}
\end{equation}
\begin{equation}
T_{R/L}[su(2)_k]=\frac{1}{k+2} \sum_{a=1}^3 J_{R/L}^aJ_{R/L}^a,
\label{2.9}
\end{equation}
and
\begin{equation}
T_{R/L}[su(k)_2]=\frac{1}{2+k} \sum_{A=1}^{k^2-1} J_{R/L}^A J_{R/L}^A,
\label{2.10}
\end{equation}
with the following currents:
\begin{equation}
J_{R/L}= \sum_{\sigma=1}^{2}\sum_{i=1}^{k} \psi^{\ast}_{R/L,\sigma,i} \psi_{R/L,\sigma,i},
\label{2.11}
\end{equation}
\begin{equation}
J_{R/L}^a=\sum_{\sigma,\rho=1}^2 \sum_{i=1}^{k} \psi^{\ast}_{R/L,\sigma,i}\frac{\sigma_{\sigma\rho}^a}{2} \psi_{R/L,\rho,i}
\label{2.12}
\end{equation}
and 
\begin{equation}
J_{R/L}^A=\sum_{\sigma=1}^2\sum_{i,j=1}^k\psi^{\ast}_{R/L,\sigma,i}T_{ij}^A\psi_{R/L,\sigma,j}.
\label{2.13}
\end{equation}
In the above expressions we are omitting the bundle index $m$. Moreover, $\sigma^a/2$ and $T^{A}$ are the 
generators of $SU(2)$ and $SU(k)$, respectively. 

The next step is to introduce interactions between wires inside a bundle in order to gap specific sectors of the theory.
As shown in \cite{Huang}, by means of the computation of renormalization group beta functions, the following interactions render a gap for different sectors inside a bundle. 
The $U(1)$ interaction corresponding to a generalized Umklapp process,
\begin{equation}
\mathcal{L}_{u(1)_k}=- g_{u(1)} \left(\prod_{i=1}^k \prod_{\alpha=1}^2\psi^{\ast}_{R,\alpha,i} \right) 
\left(\prod_{i=k}^1 \prod_{\alpha=2}^1\psi_{L,\alpha,i} \right) + [\psi_R^{\ast}\rightarrow \psi_L, \psi_L\rightarrow \psi_R^{\ast}],
\label{2.14}
\end{equation}
gaps the $U(1)$ charge sector of wires $1,...,k$ inside a bundle, just as in the case of Luttinger liquid \cite{Affleck}.
Gapping the charge sector means that we are describing non-Abelian spin liquids, not quantum Hall states. 
The $SU(2)$ current-current interaction,
\begin{equation}
\mathcal{L}_{su(2)}=-\lambda_{SU(2)}\sum_{a=1}^3 J_R^a J_L^a,
\label{2.15}
\end{equation}
gaps the $SU(2)$ sector for the wires $1,...,k$ for $\lambda_{SU(2)}>0$.
The $SU(k)$ current-current interaction,
\begin{equation}
\mathcal{L}_{su(k)}=-\lambda_{SU(k)}\sum_{A=1}^{k^2-1} J_R^A J_L^A,
\label{2.16}
\end{equation}
gaps the $SU(k)$ sector for the wires $1,...,k$ for $\lambda_{SU(k)}>0$.

In order to produce interesting non-Abelian topological phases,  
we choose to gap firstly the $U(1)$ and $SU(k)$ sectors. 
The remaining sector is gapless and supports a $SU(2)\times SU (2)$ current algebra, 
\begin{equation}
g_{k,k'}\equiv \bigoplus_{m=1}^{\mathcal{N}}su(2)_{k}\oplus su(2)_{k'}.
\label{2.17}
\end{equation}
Thus the low energy theory is characterized by a central charge
\begin{eqnarray}
c[g_{k,k'}]&=& \sum_{m=1}^{\mathcal{N}}\left(c[su(2)_k]+c[su(2)_{k'}]\right)\nonumber\\
&=& \mathcal{N} \left(\frac{3k}{k+2}+\frac{3k'}{k'+2}\right).
\label{2.18}
\end{eqnarray}
Note that this is just a number $\mathcal{N}$ of copies of a theory with central charge equal to the term between brackets.
This is so because so far we have considered only intra-bundle interactions. Thus it still does not constitute
a two dimensional phase. To this end we need to introduce inter-bundle interactions. 
This can be done by means of the diagonal subgroup $SU(2)$, with the corresponding Lie algebra  
\begin{equation}
h_{k,k'}\equiv \bigoplus_{m=1}^{\mathcal{N}}su(2)_{k+k'}.
\label{2.19}
\end{equation}
The diagonal generators $K_{R/L}^{a,m}$ are immediately  constructed from the sum of two copies of (\ref{2.12}):
\begin{equation}
K_{R/L}^{a,m}=J_{R/L}^{a,m}+J_{R/L}^{\prime a,m},
\label{2.20}
\end{equation}
where we have inserted the bundle index $m$.
The interactions that gap the desired conformal degrees of freedom are
\begin{equation}
\mathcal{L}_{su(2)}^{inter}=-\sum_{m=1}^{\mathcal{N}-1}\sum_{a=1}^{3}
\left( \lambda_{m}^a J_{L}^{a,m}J_{R}^{a,m+1}+\lambda_{m}^{\prime a} J_{L}^{\prime a,m}J_{R}^{\prime a,m+1}\right)
-\sum_{m=1}^{\mathcal{N}}\sum_{a=1}^3 g_ m^a K_{L}^{a,m} K_{R}^{a,m}.
\label{2.21}
\end{equation}
%Note that as the interaction Lagrangian is not invariant under time reversal symmetry the edge states will be chiral. 
As it is argued in \cite{Huang}, these interactions are conjectured to stabilize a strongly interacting fixed point by 
gaping all modes involved in the sum such that 
the remaining conformal degrees of freedom are that ones associated to the cosets
\begin{equation}
\left(\frac{su(2)_k\oplus su(2)_{k'}}{su(2)_{k+k'}}\right)_R~~~\text{and}~~~\left(\frac{su(2)_k\oplus su(2)_{k'}}{su(2)_{k+k'}}\right)_L,
\label{2.22}
\end{equation}
living in the first and last bundles, respectively. Each one of these sectors realizes a chiral conformal field theory
with central charge
\begin{eqnarray}
c(g/h)&=&3\left(\frac{k}{k+2}+\frac{k'}{k'+2}\right)-3 \frac{k+k'}{k+k'+2}\nonumber\\
&=&1-\frac{6 k'}{(k+2)(k+k'+2)}+\frac{2(k'-1)}{k'+2}.
\label{2.23}
\end{eqnarray} 
The presence of only one chirality in the opposite bundles is  due to the time reversal symmetry breaking 
introduced by the interactions in (\ref{2.21}). In the next section we shall return to the interaction Lagrangian (\ref{2.21}) 
to discuss some issues concerning its strong coupling limit. 

The central charge (\ref{2.23}) incorporates several classes of unitary conformal field theories. 
In particular, for $k'=1$, we obtain the series of minimal models
\begin{equation}
c(k)=1-\frac{6}{(k+2)(k+3)}, ~~~k=1,2,3,\ldots.
\label{2.24}
\end{equation} 
For $k'=2$, we obtain the series of superconformal minimal models 
\begin{equation}
c(k)=\frac{3}{2}\left[1-\frac{8}{(k+2)(k+4)}\right],~~~k=1,2,3,\ldots,
\label{2.25}
\end{equation}
that in addition to the conformal invariance exhibit supersymmetry. It is interesting to note that 
this is an emergent supersymmetry at the edge since the bulk is not supersymmetric. Along this line, 
the work of Ref. \cite{Grover} reported the emergence of spacetime supersymmetry at the  
boundary of a 2+1 dimensional topological superconductor, with central charge $c=7/10$. This conformal 
field theory describes the two dimensional tricritical Ising model at its critical point \cite{Qiu}.   
In the above construction, this supersymmetric edge state appears as the first member of the series (\ref{2.25}), 
which also coincides with the second member of (\ref{2.24}).

%%%%%%%%%%%%%%%%%%%%%%%%%%%%%%%%%%%%%%%%%%%%%%%%%%%%%%

\begin{figure}[!ht]
\centering
\includegraphics[scale=0.9]{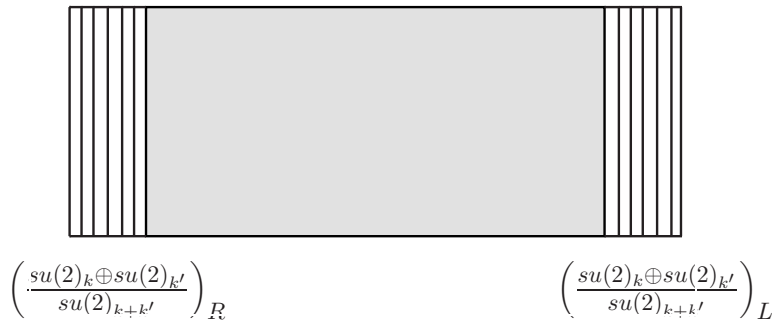}
\caption{The resulting topological phase from coupling quantum wires. The shaded region corresponds to the gapped bulk 
due to the interactions while 
the opposite ends correspond to the gapless edge states of different chiralities. }
\label{edgecoset}
\end{figure}
%%%%%%%%%%%%%%%%%%%%%%%%%%%%%%%%%%%%%%%%%%%%%%%%%%
\section{Edge Theory}\label{Sec2}

The wire construction has raised up the possibility of getting a large class of 
two-dimensional gapped topological phases of non-Abelian spin liquids with edge states 
characterized by the central charge (\ref{2.23}). The goal of this 
section is to identify the conformal field theories of the edges. We argue in favor of the plausibility in trading the strong coupling limit of the interactions among neighboring wires by direct constraints on the currents. 
This is an useful guide to the construction of the bulk theory, since we can connect the physics of the bulk and the 
edge by means of the bulk-edge correspondence, as we shall discuss in the later sections. 
The question is how to incorporate the effects of interactions into the edge theory.  

In terms of the building blocks, namely fermions propagating along the wires inside the bundles, the edge theory is 
characterized by chiral gapless degrees of freedom propagating in the first and in the last bundle, as depicted in Fig. \ref{edgecoset}.
However, there is no one-to-one correspondence between the number of wires ($k+k'$) inside those bundles 
and the number of conformal degrees of freedom. There is a fractionalization of degrees of freedom, 
such that part of them is gapped out due to the interactions and part of them remains gapless, associated to the 
coset $\frac{su(2)_k\oplus su(2)_{k'}}{su(2)_{k+k'}}$. In the following sections we will discuss how this works. 
Our strategy is to isolate the gapless sector and construct the underlying conformal field theory. In doing so we keep 
both chiralities to avoid issues with gauge anomalies \cite{Tye}. As we shall see, the final picture involves a chiral splitting
with each chiral conformal field theory living at the opposite edges of the bulk.

%%%%%%%%%%%%%%%%%%%%%%%%%%%%%%%%%%%%%%%%%%%%%%%%%%%%%%%%
\subsection{Strongly Coupled Limit}

The quantum theory for the whole system can be defined by the partition function
\begin{equation}
\mathcal{Z}=\int{\mathcal{D}\psi_R\mathcal{D}\psi_L\exp\mathrm{i}\int dx dt \left( \mathcal{L}_0+\mathcal{L}_{int}\right)},
\label{2.252}
\end{equation}
where $\mathcal{D}\psi_R\mathcal{D}\psi_L$ is a shorthand for the measure
\begin{equation}
\mathcal{D}\psi_R\mathcal{D}\psi_L=\prod_{m=1}^{\mathcal{N}}\prod_{i=1}^{k}\prod_{i'=1}^{k'}\prod_{\sigma=1}^2\mathcal{D}\psi_{R,\sigma,i}^m\mathcal{D}\psi_{L,\sigma,i}^m\mathcal{D}\psi_{R,\sigma,i'}^{m}\mathcal{D}\psi_{L,\sigma,i'}^{m}\mathcal{D}\psi^{m\ast}_{R,\sigma,i}\mathcal{D}\psi^{m\ast}_{L,\sigma,i}\mathcal{D}\psi^{m\ast}_{R,\sigma,i'}\mathcal{D}\psi^{m\ast}_{L,\sigma,i'}.
\end{equation}
The free Lagrangian $\mathcal{L}_0$ is given by (\ref{2.1a}) and the interacting Lagrangian $\mathcal{L}_{int}$ is the sum
\begin{equation}
\mathcal{L}_{int}=\sum^{\mathcal{N}}_{m=1}\left(\mathcal{L}_{u(1)_k}^m+\mathcal{L}_{u(1)_{k'}}^m+\mathcal{L}^m_{su(k)}+\mathcal{L}^m_{su(k')}\right)+\mathcal{L}_{su(2)}^{inter},
\end{equation}
with $\mathcal{L}_{u(1)_k}^m$ and $\mathcal{L}^m_{su(k)}$ given by (\ref{2.14}) and (\ref{2.16}), respectively, 
with similar expressions for $k\rightarrow k'$, and the interaction $\mathcal{L}_{su(2)}^{inter}$ given by (\ref{2.21}).

The fact that the interactions among the quantum wires provide a mass gap to the interacting modes enables us to describe the low-energy strongly coupled effective theory, where the gapped modes are projected out, in terms of a set of constraints on an initially free fermion model. As we will show in detail in the next section, to suppress the correct gapped modes, we can impose constraints on the currents that partake of the interactions among the quantum wires. We then claim that the conjectured topological phase obtained by the quantum wires construction can be realized in terms of the initial set of free fermions propagating inside the wires but subjected to constraints of the form $current=0$ on all the initially interacting currents.

The argument that this constrained system is the strong coupling limit of the interacting quantum wires picture relies on the validity of the conjecture about the stability of the strongly coupled phase. The subtle point of this conjecture concerns the non-commutativity of the two kinds of terms in the interacting Lagrangian (\ref{2.21}) that partake of the interactions between neighboring bundles, $JJ$-terms, with those that gap the diagonal subgroup inside each bundle, $KK$-terms. In the strong coupled limit the competition between these interactions might not result in the expected topological phase since the currents are not separately conserved.

%%%%%%%%%%%%%%%%%%%%%%%%%%%%%%%%%%%%%%%%%%%%%%%%%%%%%%%%%%%%%%%

The perturbative analysis of the renormalization group shows that coupling constants $\lambda_m$, $\lambda_m'$ and $g_m$ flow independently to the strong coupled limit if they are initially positive \cite{Huang}. 
While this is compatible with the general picture expected to get a two dimensional topological phase, it does not mean that 
we can reach the desired topological phase since the inter and intra bundle interactions compete with each other.  Outside the perturbative regime we do not have control on the coupling constants and it is a formidable task to get a definitive conclusion. 
To ultimately decide if the topological phase can be stabilized involves to access nonperturbative information, which is beyond the scope of this work.
We can, however, consider certain limiting cases to acquire some intuition on what physics we can expect\footnote{We would like to thank Claudio Chamon and Christopher Mudry to call our attention to this point.}.  To appreciate this point
we will focus just on the interacting Lagrangian containing non-commuting terms (\ref{2.21}). Let us rewrite it 
by using auxiliary fields and discuss the problem in terms of the equivalent Lagrangian
\begin{eqnarray}
\mathcal{L}_{su(2)}^{inter}&=&\sum^{\mathcal{N}-1}_{m=1}\left(C^{m}_-J_{L}^{m}+C^{m+1}_+J_{R}^{m+1}+\frac{1}{\lambda_m}C^{m+1}_+C^{m}_-\right.\nonumber\\
&+&\left.C^{\prime m}_-J_{L}^{\prime m}+C^{\prime m+1}_+J_{R}^{\prime m+1}+\frac{1}{\lambda'_m}C^{\prime m+1}_+C^{\prime m}_-\right)\nonumber\\
&+&\sum^{\mathcal{N}}_{m=1}\left(D^{m}_-K_{L}^{m}+D^{m}_+K_{R}^{m}+\frac{1}{g_m}D^{m}_-D^{m}_+\right).
\label{2.251}
\end{eqnarray}
The equivalence of the Lagrangians (\ref{2.21}) and (\ref{2.251}) is achieved by using the equations of motion for the auxiliary fields in the Lagrangian (\ref{2.251}). We can then verify that the original interactions are restored.

The two limiting cases we will consider are 
\begin{equation}
\frac{\lambda}{g}\,\,\,\rightarrow\,\,\,\left\{
\begin{array}
[c]{cc}%
\gg 1&   \\
%\sim 1 & \\
\ll 1 &
\end{array}
\right..
\end{equation}
We will analyze the resulting phases in these different situations.

%%%%%%%%%%%%%%%%%%%%%%%%%%%%%%%%%%%%%%%%%%%%%%%%%%%
\subsubsection{Case $\lambda/g \gg 1$}

In this case the mixing terms in the first and second lines of (\ref{2.251}) are negligible and the integrals over the gauge fields implement the 
constraints
\begin{equation}
J_L^m=J_{L}^{\prime m}=0 ~~~\text{and}~~~J_R^{m+1}=J_{R}^{\prime m+1}=0,
\label{7.1}
\end{equation}
with $m=1,...,\mathcal{N}-1$. Thus, for $g_m\rightarrow 0$, the constraints (\ref{7.1}) render a topological phase with boundary gapless modes supporting the algebra
\begin{equation}
\left(su(2)_k\oplus su(2)_{k'}\right)_R~~~\text{and}~~~\left(su(2)_k\oplus su(2)_{k'}\right)_L.
\end{equation}
Since $m=1,\ldots,\mathcal{N}-1$, the only gapless sectors correspond to the right modes in the first bundle and the left modes in the last one.
For non-vanishing intra-bundle interactions, $g_m\neq 0$, the constraints (\ref{7.1}) imply $K_L^m=0$ and $K_R^{m+1}=0$, such that the remaining terms of 
(\ref{2.251}) are
\begin{equation}
\mathcal{L}_{su(2)}^{inter}\sim \left(D_{+}^1 K_R^1+\frac{1}{g} D_{-}^1D_{+}^1\right)+\sum_{m=2}^{\mathcal{N}-1}\frac{1}{g} D_{-}^m D_{+}^m+
 \left(D_{-}^{\mathcal{N}} K_L^{\mathcal{N}}+\frac{1}{g} D_{-}^{\mathcal{N}}D_{+}^{\mathcal{N}}\right).
 \label{7.2}
\end{equation}
To understand the role that the intra-bundle interactions play in this phase, we can proceed, for example, by integrating over the fields $D_{-}^m$, $m=1,...,\mathcal{N}$. This will not lead to any additional constraint over the bundles. Only the constraints in (\ref{7.1}) play
a role, by fully gapping the bulk. This phase is then stable under small perturbations of the intra-bundle interactions, leaving chiral edge states supporting the algebra 
\begin{equation}
\left(su(2)_k\oplus su(2)_{k'}\right)_R~~~\text{and}~~~\left(su(2)_k\oplus su(2)_{k'}\right)_L,\label{5}
\end{equation}
which correspond to a non-coset topological phase. Of course, the same conclusions are obtained if we integrate out the fields $D_{+}^m$.

%%%%%%%%%%%%%%%%%%%%%%%%%%%%%%%%%%%%%%%%%%%%%%%%
\subsubsection{Case $\lambda/g \ll 1$}

In this limit the mixing term $\frac{1}{g_m}D^m_-D^m_+$ in (\ref{2.251})  becomes negligible compared to the remaining ones. Then the last line effectively imposes the constraints $\delta(K^m_L)$ and $\delta(K^m_R)$, which is equivalent to impose
\begin{equation}
J^m_L=-J_L^{m\prime},\ \ \ , J^m_R=-J_R^{m\prime}.\label{2.3z}
\end{equation}
For vanishing inter bundle interactions, $\lambda_m\rightarrow 0$, these constraints generate a set of uncoupled one-dimensional conformal field theories
corresponding to the coset $su(2)_k\oplus su(2)_{k^{\prime}}/su(2)_{k+k^{\prime}}$ for the left and right modes. 
This gapless phase is not stable under small pertubations of inter-bundle interactions. 

For non-vanishing $\lambda_m$ the constraints (\ref{2.3z}) turn the Lagrangian (\ref{2.251}) into
\begin{eqnarray}
\mathcal{L}_{su(2)}^{inter}&=&\sum^{\mathcal{N}-1}_{m=1}\left(C^{m}_-J_{L}^{m}+C^{m+1}_+J_{R}^{m+1}+\frac{1}{\lambda_m}C^{m+1}_+C^{m}_-\right.\nonumber\\
&-&\left.C^{\prime m}_-J_{L}^{m}-C^{\prime m+1}_+J_{R}^{m+1}+\frac{1}{\lambda'_m}C^{\prime m+1}_+C^{\prime m}_-\right).
\label{2.255}
\end{eqnarray}
After integrating the auxiliary fields we get
\begin{eqnarray}
\mathcal{L}_{su(2)}^{inter}&=&-\sum^{\mathcal{N}-1}_{m=1}(\lambda_m+\lambda^{\prime}_m)J_{L}^{m}J_R^{m+1},
\label{2.256}
\end{eqnarray}
which is exactly the inter bundle interactions. The difference, however, is that now the flow of these constants is determined by 
the renormalization group analysis in the non perturbative region with large $g$.
Despite the difficulty of doing that, three scenarios rise up:
 
i) the constant couplings are marginal irrelevant. In this case, as we go to low energy we get back the initial gapless phase;  

ii)  the coupling constants are marginally exact. This will not generate a gap and the theory remains critical; 

iii) the coupling constants are marginal relevant. In this case, the above interactions destabilize the initial gapless phase, leading the system to a gapped topological phase. Since they couple the gapless modes between neighboring bundles, the only decoupled modes are the right modes in the first bundle and the left modes in the last one. This will delivery a strongly coupled phase supporting chiral edge states associated to the coset structure
\begin{equation}
\left(\frac{su(2)_k\oplus su(2)_{k'}}{su(2)_{k+k'}}\right)_R~~~\text{and}~~~\left(\frac{su(2)_k\oplus su(2)_{k'}}{su(2)_{k+k'}}\right)_L,
\end{equation}
which is a non-Abelian topological phase. The best we can do now is to assume this case and to explore some interesting consequences. 
This analysis will be the object of the remaining of the paper. As we shall discuss in detail, in this limit the interactions 
can be represented in terms of constraints on currents. Among other things, this enable us to identify the edge states directly in terms of the original fermions and then to use the bulk-edge correspondence to construct a low energy bulk theory for that class of topological phases.

\begin{figure}[!h]
\centering
\includegraphics[scale=0.7]{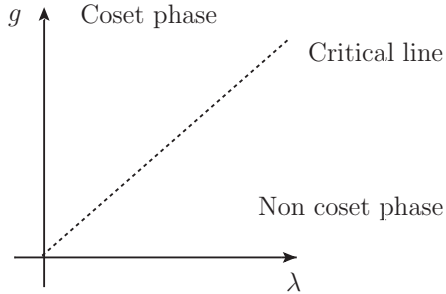}
\caption{Sketch of phase diagram compatible with scenario iii). There is a critical line with a gap closing separating two distinct topological phases.}
\label{PhaseDiag}
\end{figure}

In passing from the limiting case  $\lambda/g \gg 1$ to $\lambda/g \ll 1$ we find out two distinct strongly coupled topological
phases, indicating that a phase transition with gap closing must take place in the non perturbative region where the inter and intra bundle interactions compete to each other. A qualitative phase diagram is depicted in Fig. \ref{PhaseDiag}. 
This general pattern is explicitly confirmed in the work of Ref. \cite{Tsvelik}, where it is analyzed a mean field solution for a system of quantum wires built up from Majorana fermions with interactions sharing some features of that ones in (\ref{2.21}). 

To proceed we take the strong coupling limit,  $(\lambda,g)\rightarrow\infty$, understanding that 
we are running in the region comprised in the case iii) where the coset topological phase is realized, without crossing the critical line.  
We can do this directly in the Lagrangian (\ref{2.21}).
In this limit the quadratic terms in the auxiliary fields are suppressed and we get a set of Lagrange multiplier fields enforcing the constraints $current=0$. In the partition function these constraints are implemented through delta functions $\delta(current)$. The issue of having non-commuting terms in the Lagrangian (\ref{2.251}) can be identified in the strong coupling limit with overlapping delta functions. In fact, for all the bundles inside the bulk, $m=2,\ldots,\mathcal{N}-1$, the constraints $\delta(J^m)$ and $\delta(J^{\prime m})$ already integrate out the associated degrees of freedom. The extra constraints $K^m_{R/L}=0$ are redundant for these bundles. This problem could be avoided by just turning off the $K$-interactions without affecting the strongly coupled topological phase. However, for the first and last bundles, $m=1$ and $m=\mathcal{N}$, respectively, we do not have this option, since the $K$-coupling is essential for gapping the right number of modes inside the edge bundles 
in order to produce a non-Abelian topological phase. For those bundles, only the $K^1_L$ and $K^{\mathcal{N}}_R$ constraints are redundant, and as said this redundancy cannot be avoided by turning off the $K$-coupling without affecting the expected topological phase.

%%%%%%%%%%%%%%%%%%%%%%%%%%%%%%%%%%%%%%%%%%%%%%%%%%%%%%%%%%%

By taking $(\lambda,g)\rightarrow\infty$ in (\ref{2.251}) we obtain
\begin{eqnarray}
\mathcal{Z}&=&\int\mathcal{D}\psi_R\mathcal{D}\psi_L\underbrace{\delta(J_{u(1)})\delta(J_{su(k)})\delta(J_{u'(1)})\delta(J_{su(k')})}_{\text{all bundles}}
\underbrace{ \delta(J_{su(2)_k}) \delta(J_{su(2)_{k'}}) \delta(K_{su(2)_{k+k'}}) }_{\text{bulk}}
\nonumber\\
&\times& \delta(J^{a,1}_L)\delta(J^{\prime a,1}_L)\delta(K^{a,1}_R)\delta(K^{a,1}_L)\delta(J^{a,\mathcal{N}}_R)\delta(J^{\prime a,\mathcal{N}}_R)\delta(K^{a,\mathcal{N}}_L)\delta(K^{\mathcal{N}}_R)
\exp \mathrm{i} \int dtdx\mathcal{L}_0.
\label{2.253}
\end{eqnarray}
The delta functions in the first line compactly represent the constraints in the $U(1)$, $U'(1)$, $SU(k)$,  and $SU(k')$ sectors in all bundles of the system, in addition to the constraints in the sectors $SU(2)_k$, $SU(2)_{k'}$, and $SU(2)_{k+k'}$ for the bundles of the bulk, i.e., 
excluding the bundles of the edges (first and last), which are explicitly written in the second line since they need a more detailed analysis.

The constraints for the $SU(2)$-currents $J^1_L$ and $J^{\prime 1}_L$ make the constraint in $K^1_L$ to be redundant. The same happens to $K^{\mathcal{N}}_R$ due to constraints in $J^{\mathcal{N}}_R$ and $J^{\prime \mathcal{N}}_R$ and to all the diagonal constraints in $K^m_L$ and $K^m_R$ inside the bulk due to the delta functions $\delta(J^m_{R/L})$ and $\delta(J^{\prime m}_{R/L})$, with $m=2,\ldots,\mathcal{N}-1$. All these redundant delta functions for the diagonal currents contribute with factor of $\delta(0)$ to the partition function and render it ill-defined in the strong coupling limit. However, for each one of these diagonal currents we can understand the singular deltas as functional integrals over the accompanying respective auxiliary fields in (\ref{2.252}). Since these fields decouple in this limit, we can simply absorb the infinite factors into the functional measure without prejudice to the low-energy topological phase. Of course, correlations functions of the theory are insensitive to this.

As we shall see, the role of the constraints is to delete from the spectrum the modes in the representation of group related to the respective current. For the bundles in the bulk, the constraints are sufficient to suppress all the modes. We will end up with chiral gapless excitations living on both edges of the system describing a conformal field theory with central charge given by (\ref{2.23}).

With all these qualifications we can write the effective partition function compactly as
\begin{eqnarray}
\mathcal{Z}&=&\mathcal{Z}_{bulk}\int\mathcal{D}\psi_R\mathcal{D}\psi_L\bar{\delta}(1)\bar{\delta}(\mathcal{N})\delta(J^{a,1}_L)\delta(J^{a',1}_L)\delta(K^{a,1}_R)\delta(J^{a,\mathcal{N}}_R)\delta(J^{a',\mathcal{N}}_R)\delta(K^{a,\mathcal{N}}_L)\nonumber\\
&\times&\exp \mathrm{i}\int{dtdx\mathcal{L}(1,\mathcal{N})},
\label{2.253a}
\end{eqnarray}
with 
\begin{equation}
\bar{\delta}(1)\equiv\underbrace{\delta(J^{1}_{R/L})\delta(J^{\prime 1}_{R/L})}_{u(1)_k,~u(1)_{k'}}\underbrace{\delta(J^{A,1}_{R/L})\delta(J^{A',1}_{R/L})}_{su(k),~su(k')}~~~\text{and}~~~\bar{\delta}(\mathcal{N}) \equiv \underbrace{\delta(J^{\mathcal{N}}_{R/L})\delta(J^{\prime \mathcal{N}}_{R/L})}_{u(1)_k,~u(1)_{k'}}\underbrace{\delta(J^{A,\mathcal{N}}_{R/L})\delta(J^{A',\mathcal{N}}_{R/L})}_{su(k),~su(k')},
\end{equation}
remembering that  $A=1,\ldots,k^2-1$, $A'=1,\ldots,k^{\prime 2}-1$, and $a,a'=1,2,3$. 
The Lagrangian $\mathcal{L}(1,\mathcal{N})$ as well as the functional measure refer only to the fermions of the first and last bundles.  
The decoupled factor $\mathcal{Z}_{bulk}$, which is fully gapped, can be absorbed into the measure.
In the next section we will start a detailed analysis of the partition function (\ref{2.253a}) in order to 
unveil its gapless content.  

This whole picture makes it plausible to describe the non-Abelian topological phase in terms of constrained fermions 
that, as we shall see, is reinforced by the 
equality of central charges of the edges in both approaches.

%%%%%%%%%%%%%%%%%%%%%%%%%%%%%%%%%%%%%%%%%%%%%%%%%%%%%%%%%%%%%%%%%
\subsection{Constrained Fermions at the Edges}

According to the above discussion, the dismembering of degrees of freedom due to interactions can be incorporated into 
the edge bundles in terms of constraints over currents. In an operator formalism, this is equivalent to imposing
selection rules for gapless physical states. For completeness,
we write down explicitly all selection rules contained in partition function (\ref{2.253a}). 
For the first bundle, we have the following selection rules:
\begin{equation}
J^{1}_{R} | \text{phys}\rangle =0~~~\text{and}~~~J^{1}_L | \text{phys}\rangle =0,~~~\text{for sector}~~u(1)_k;
\label{3.1a}
\end{equation}
\begin{equation}
J^{\prime 1}_R | \text{phys}\rangle =0~~~\text{and}~~~J^{\prime 1}_L | \text{phys}\rangle =0,~~~\text{for sector}~~u(1)_{k'};
\label{3.1b}
\end{equation}
\begin{equation}
J^{A,1}_R | \text{phys}\rangle =0~~~\text{and}~~~J^{A,1}_L | \text{phys}\rangle =0,~~~\text{for sector}~~su(k)_2;
\label{3.1c}
\end{equation}
\begin{equation}
J^{\prime A,1}_R | \text{phys}\rangle =0~~~\text{and}~~~J^{\prime A,1}_L | \text{phys}\rangle =0,~~~\text{for sector}~~su(k')_2;
\label{3.1d}
\end{equation}
\begin{equation}
J^{a,1}_L | \text{phys}\rangle=0,  ~~~\text{for sector}~~su(2)_k;
\label{3.1e}
\end{equation}
\begin{equation}
J^{\prime a,1}_L | \text{phys}\rangle=0,~~~\text{for sector}~~su(2)_{k'};
\label{3.1f}
\end{equation}
\begin{equation}
K^{a,1}_R | \text{phys}\rangle =0,~~~\text{for sector}~~su(2)_{k+k'}.
\label{3.1g}
\end{equation}
Notice that we have also included constraints for sectors $u(1)_k$, $u(1)_{k'}$, $su(k)$ and $su(k')$, that 
can be handled in a similar way, but involving only intra-bundle interactions. 
Although the $U(1)$ charge-sector interaction is not of the current-current form, the constraint on the $U(1)$ current is equivalent in the low-energy limit to integrate gapped degrees of freedom out of the spectrum.
We have a similar structure in the last bundle, where the rules (\ref{3.1e}) and (\ref{3.1f}) are 
replaced by the corresponding right currents, while 
(\ref{3.1g}) is replaced by the corresponding left current,
\begin{equation}
J^{\mathcal{N}}_{R} | \text{phys}\rangle =0~~~\text{and}~~~J^{\mathcal{N}}_L | \text{phys}\rangle =0,~~~\text{for sector}~~u(1)_k;
\label{3.1aa}
\end{equation}
\begin{equation}
J^{\prime \mathcal{N}}_R | \text{phys}\rangle =0~~~\text{and}~~~J^{\prime \mathcal{N}}_L | \text{phys}\rangle =0,~~~\text{for sector}~~u(1)_{k'};
\label{3.1bb}
\end{equation}
\begin{equation}
J^{A,\mathcal{N}}_R | \text{phys}\rangle =0~~~\text{and}~~~J^{A,\mathcal{N}}_L | \text{phys}\rangle =0,~~~\text{for sector}~~su(k)_2;
\label{3.1cc}
\end{equation}
\begin{equation}
J^{\prime A,\mathcal{N}}_R | \text{phys}\rangle =0~~~\text{and}~~~J^{\prime A,\mathcal{N}}_L | \text{phys}\rangle =0,~~~\text{for sector}~~su(k')_2;
\label{3.1dd}
\end{equation}
\begin{equation}
J^{a,\mathcal{N}}_R | \text{phys}\rangle=0,  ~~~\text{for sector}~~su(2)_k;
\label{3.1ee}
\end{equation}
\begin{equation}
J^{\prime a,\mathcal{N}}_R | \text{phys}\rangle=0,~~~\text{for sector}~~su(2)_{k'};
\label{3.1ff}
\end{equation}
\begin{equation}
K^{a,\mathcal{N}}_L | \text{phys}\rangle =0,~~~\text{for sector}~~su(2)_{k+k'}.
\label{3.1gg}
\end{equation}  
In these expressions the state $| \text{phys}\rangle$ represent the physical conformal degrees of freedom of the first and last bundles, 
$| \text{phys}\rangle\equiv | R\rangle\otimes |L\rangle $.

Now we go back to the partition function (\ref{2.253a}), which takes into account all of these conditions. 
We write it again for convenience, 
\begin{eqnarray}
\mathcal{Z}&=&\int \mathcal{D}\psi_R \mathcal{D}\psi_L  \delta(1) \delta(\mathcal{N})
\exp \mathrm{i} \int dt dx\, \mathcal{L}.
\label{3.2}
\end{eqnarray}
This partition function is the compact form for
\begin{eqnarray}
\delta(1)&\equiv& \delta(J_R^1 )\delta(J_L^1)
 \delta(J^{\prime 1}_R )\delta(J^{\prime 1}_L)
 \delta(J^{A,1}_R)\delta(J^{A,1}_L)
\delta(J^{\prime A,1}_R)\delta(J^{\prime A,1}_L)\nonumber\\
&\times&\delta (J^{a,1}_L )\delta (J^{\prime a,1}_L ) \delta (K^{a,1}_R )
\label{3.2a}
\end{eqnarray}
and
\begin{eqnarray}
\delta(\mathcal{N})&\equiv&\delta(J_R^{\mathcal{N}} )\delta(J_L^{\mathcal{N}})
 \delta(J^{\prime \mathcal{N}}_R )\delta(J^{\prime \mathcal{N}}_L)
 \delta(J^{A,\mathcal{N}}_R)\delta(J^{A,\mathcal{N}}_L)
\delta(J^{\prime A,\mathcal{N}}_R)\delta(J^{\prime A,\mathcal{N}}_L)\nonumber\\
&\times&
\delta (J^{a,\mathcal{N}}_R )\delta (J^{\prime a,\mathcal{N}}_R ) \delta (K^{a,\mathcal{N}}_L ),
\label{3.2b}
\end{eqnarray}
with the functional measure defined as
\begin{eqnarray}
\mathcal{D}\psi_R \mathcal{D}\psi_L&\equiv& \prod_{\sigma=1}^2\prod_{i=1}^k \prod_{i=1}^{k'}\mathcal{D}\psi_{R,\sigma,i}^{1\ast}
\mathcal{D}\psi_{R,\sigma,i}^1 \mathcal{D}\psi_{R,\sigma,i'}^{1\ast}\mathcal{D}\psi_{R,\sigma,i'}^1 \nonumber\\
&\times& \prod_{\sigma=1}^2\prod_{i=1}^k \prod_{i=1}^{k'}\mathcal{D}\psi_{L,\sigma,i}^{1\ast}
\mathcal{D}\psi_{L,\sigma,i}^1 \mathcal{D}\psi_{L,\sigma,i'}^{1\ast}\mathcal{D}\psi_{L,\sigma,i'}^1\nonumber\\
&\times&
\prod_{\sigma=1}^2\prod_{i=1}^k \prod_{i=1}^{k'}\mathcal{D}\psi_{R,\sigma,i}^{\mathcal{N}\ast}
\mathcal{D}\psi_{R,\sigma,i}^{\mathcal{N}} \mathcal{D}\psi_{R,\sigma,i'}^{\mathcal{N}\ast}\mathcal{D}
\psi_{R,\sigma,i'}^{\mathcal{N}} \nonumber\\
&\times& \prod_{\sigma=1}^2\prod_{i=1}^k \prod_{i=1}^{k'}\mathcal{D}\psi_{L,\sigma,i}^{\mathcal{N}\ast}
\mathcal{D}\psi_{L,\sigma,i}^{\mathcal{N}} \mathcal{D}\psi_{L,\sigma,i'}^{\mathcal{N}\ast}
\mathcal{D}\psi_{L,\sigma,i'}^{\mathcal{N}}.
\label{3.4}
\end{eqnarray}
The Lagrangian in (\ref{3.2}) involves the contributions of all the wires in the first and last bundles
\begin{eqnarray}
\mathcal{L}&=& \sum_{\sigma=1}^2\sum_{i=1}^k \left[\mathrm{i}\psi_{R,\sigma,i}^{1\ast}(\partial_t+\partial_x)
\psi_{R,\sigma,i}^1 
+\mathrm{i}\psi_{L,\sigma,i}^{1\ast}(\partial_t-\partial_x)\psi_{L,\sigma,i}^1\right]\nonumber\\
&+& \sum_{\sigma=1}^2\sum_{i'=1}^{k'}\left[\mathrm{i}\psi_{R,\sigma,i'}^{1\ast}(\partial_t+\partial_x)\psi_{R,\sigma,i'}^1
+\mathrm{i}\psi_{L,\sigma,i'}^{1\ast}(\partial_t-\partial_x)\psi_{L,\sigma,i'}^1\right]\nonumber\\
&+&\sum_{\sigma=1}^2\sum_{i=1}^k \left[\mathrm{i}\psi_{R,\sigma,i}^{\mathcal{N}\ast}(\partial_t+\partial_x)
\psi_{R,\sigma,i}^{\mathcal{N}} 
+\mathrm{i}\psi_{L,\sigma,i}^{\mathcal{N}\ast}(\partial_t-\partial_x)\psi_{L,\sigma,i}^{\mathcal{N}}\right]\nonumber\\
&+& \sum_{\sigma=1}^2\sum_{i'=1}^{k'}\left[\mathrm{i}\psi_{R,\sigma,i'}^{\mathcal{N}\ast}(\partial_t+\partial_x)\psi_{R,\sigma,i'}^{\mathcal{N}}
+\mathrm{i}\psi_{L,\sigma,i'}^{\mathcal{N}\ast}(\partial_t-\partial_x)\psi_{L,\sigma,i'}^{\mathcal{N}}\right].
\label{3.3}
\end{eqnarray}
Now we write back the functional deltas in terms of Lagrange multiplier fields as it follows:
\begin{equation}
\delta (J_{R/L}^1 ) =\int \mathcal{D} A_{\pm}^1 \exp \mathrm{i} \int dt dx\, A_{\pm}^1 J_{R/L}^1 ,~~~ \text{similarly for}~~~ (1\rightarrow\mathcal{N}),
\label{3.5}
\end{equation}
\begin{equation}
\delta (J^{\prime 1}_{R/L} ) =\int \mathcal{D} A^{\prime 1}_{\pm} \exp \mathrm{i} \int dt dx\, A^{\prime 1}_{\pm} J^{\prime 1}_{R/L}, 
~~~ (1\rightarrow\mathcal{N}),
\label{3.5a}
\end{equation}
where all $A$'s are $U(1)$ Abelian gauge fields;
\begin{equation}
\delta(J^{A,1}_{R/L})=\int \mathcal{D} B_{\pm}^1 \exp \mathrm{i} \int dt dx\, B_{\pm}^{A,1}J^{A,1}_{R/L},~~~ (1\rightarrow\mathcal{N}),
\label{3.6}
\end{equation}
where $B$'s are $SU(k)$ non-Abelian gauge fields;
\begin{equation}
\delta(J^{\prime A,1}_{R/L})=\int \mathcal{D} B_{\pm}^{\prime 1} \exp \mathrm{i} \int dt dx\, B_{\pm}^{\prime A,1}J^{\prime A,1}_{R/L},~~~ (1\rightarrow\mathcal{N}),
\label{3.7}
\end{equation}
where $B'$'s are  $SU(k')$ non-Abelian gauge fields;
\begin{equation}
\delta(J_L^{a,1})=\int\mathcal{D}C_{-}^1 \exp\mathrm{i}\int dt dx \, C_{-}^{a,1} J_L^{a,1}, ~~~(1\rightarrow \mathcal{N},~L\rightarrow R,~-\rightarrow +),
\end{equation}
\begin{equation}
\delta(J_L^{\prime a,1})=\int\mathcal{D}C^{\prime 1}_{-} \exp \mathrm{i}\int dt dx \, C_{-}^{\prime a,1} J_L^{\prime a,1},~~~(1\rightarrow \mathcal{N},~L\rightarrow R,~-\rightarrow +),
\label{3.7a}
\end{equation}
\begin{equation}
\delta (K^{a,1}_{R} )=\int \mathcal{D} C^1_{+} \exp \mathrm{i} \int dt dx\, C_{+}^{a,1} K_{R}^{a,1},~~~(1\rightarrow \mathcal{N},~
R\rightarrow L,~+\rightarrow-),
\label{3.8}
\end{equation}
where all $C$'s are $SU(2)$ non-Abelian gauge field.

The partition function becomes
\begin{equation}
\mathcal{Z}=\int \mathcal{D}\psi_R \mathcal{D}\psi_L \mathcal{D}G \exp \mathrm{i} \int dt dx \,\tilde{\mathcal{L}},
\label{3.9}
\end{equation}
with the measure for the gauge fields defined as
\begin{equation}
\mathcal{D}G \equiv \mathcal{D}A_{\pm}^1 \mathcal{D}A_{\pm}^{\mathcal{N}} \mathcal{D}A^{\prime 1}_{\pm} \mathcal{D}A^{\prime \mathcal{N}}_{\pm}
\mathcal{D}B_{\pm}^1 \mathcal{D}B_{\pm}^{\mathcal{N}} \mathcal{D}B^{\prime 1}_{\pm} \mathcal{D}B^{\prime \mathcal{N}}_{\pm}
\mathcal{D}C^1_+ \mathcal{D}C^1_- \mathcal{D}C^{\prime 1}_-
\mathcal{D}C^{\mathcal{N}}_- \mathcal{D}C^{\mathcal{N}}_+ \mathcal{D}C^{\prime \mathcal{N}}_+
\end{equation}
and
\begin{eqnarray}
\tilde{\mathcal{L}}&=& 
\sum_{\sigma,\rho=1}^2\sum_{i,j=1}^k \left[\psi_{R,\sigma,i}^{1\ast}
[(\mathrm{i}\partial_+ +A_+^1)\delta_{\sigma\rho}\delta_{ij}+(B_+^1)^{ij}\delta_{\sigma\rho} +(C_+^1)^{\sigma\rho}\delta_{ij}]
\psi_{R,\rho,j}^1 \right.\nonumber\\
&+&\left.\psi_{L,\sigma,i}^{1\ast}[(\mathrm{i}\partial_{-}+A_-^1)\delta_{\sigma\rho}\delta_{ij}+
(B_-^1)^{ij}\delta_{\sigma\rho}
+(C_{-}^1)^{\sigma\rho}\delta_{ij}]\psi_{L,\rho,j}^1\right]\nonumber\\
&+& \sum_{\sigma\rho=1}^2\sum_{i',j'=1}^{k'}\left[\psi_{R,\sigma,i'}^{1\ast}[(\mathrm{i}\partial_{+}+A^{\prime 1}_+)\delta_{\sigma\rho}\delta_{i'j'}
+(B_+^{\prime 1})^{ i'j'}\delta_{\sigma\rho}+(C_+^1)^{\sigma\rho}\delta_{i'j'}]\psi_{R,\rho,j'}^1 \right.\nonumber\\
&+&\left.\psi_{L,\sigma,i'}^{1\ast}[(\mathrm{i}\partial_{-}+A^{\prime 1}_-)\delta_{\sigma\rho}\delta_{i'j'}+
(B_-^{\prime 1})^{i'j'}\delta_{\sigma\rho}+(C_-^{\prime 1})^{\sigma\rho}\delta_{i'j'}]\psi_{L,\rho,j'}^1\right]\nonumber\\
&+&
\sum_{\sigma,\rho=1}^2\sum_{i,j=1}^k \left[\psi_{R,\sigma,i}^{\mathcal{N}\ast}
[(\mathrm{i}\partial_+ +A_+^{\mathcal{N}})\delta_{\sigma\rho}\delta_{ij}+(B_+^{\mathcal{N}})^{ij}\delta_{\sigma\rho} +(C_+^{\mathcal{N}})^{\sigma\rho}\delta_{ij}]
\psi_{R,\rho,j}^{\mathcal{N}} \right.\nonumber\\
&+&\left.\psi_{L,\sigma,i}^{\mathcal{N}\ast}[(\mathrm{i}\partial_{-}+A_-^{\mathcal{N}})\delta_{\sigma\rho}\delta_{ij}+
(B_-^{\mathcal{N}})^{ij}\delta_{\sigma\rho}
+(C_{-}^{\mathcal{N}})^{\sigma\rho}\delta_{ij}]\psi_{L,\rho,j}^{\mathcal{N}}\right]\nonumber\\
&+& \sum_{\sigma\rho=1}^2\sum_{i',j'=1}^{k'}\left[\psi_{R,\sigma,i'}^{\mathcal{N}\ast}[(\mathrm{i}\partial_{+}+A^{\prime \mathcal{N}}_+)\delta_{\sigma\rho}\delta_{i'j'}
+(B_+^{\prime \mathcal{N}})^{ i'j'}\delta_{\sigma\rho}+(C_+^{\prime \mathcal{N}})^{\sigma\rho}\delta_{i'j'}]\psi_{R,\rho,j'}^{\mathcal{N}} \right.
\nonumber\\
&+&\left.\psi_{L,\sigma,i'}^{\mathcal{N}\ast}[(\mathrm{i}\partial_{-}+A^{\prime \mathcal{N}}_-)\delta_{\sigma\rho}\delta_{i'j'}+
(B_-^{\prime \mathcal{N}})^{i'j'}\delta_{\sigma\rho}+(C_-^{\mathcal{N}})^{\sigma\rho}\delta_{i'j'}]\psi_{L,\rho,j'}^{\mathcal{N}}\right],
\label{3.10}
\end{eqnarray}
where we have used the shorthand $\partial_{\pm}\equiv \partial_t \pm \partial_x$.
The non-Abelian gauge fields in this expression are contracted with the corresponding generators, i.e., 
\begin{equation}
(B_{\pm}^{\mu})^{ij}\equiv \sum_{A=1}^{k^2-1} T_{ij}^A B_{\pm}^{A,\mu}~~~\text{and}~~~
(B_{\pm}^{\prime\mu})^{ i'j'}\equiv \sum_{A=1}^{k^{\prime 2}-1} T_{i'j'}^A B_{\pm}^{\prime A,\mu},~~~\mu=0,\mathcal{N}
\label{3.11}
\end{equation}
and
\begin{equation}
(C_{\pm}^{\mu})^{\sigma\rho}\equiv \sum_{a=1}^3\frac{\sigma^a_{\rho\sigma}}{2} C_{\pm}^{a,\mu}~~~\text{and}~~~
(C_{\pm}^{\prime\mu})^{\sigma\rho}\equiv \sum_{a=1}^3\frac{\sigma^a_{\rho\sigma}}{2} C_{\pm}^{\prime a,\mu},
~~~\mu=0,\mathcal{N}.
\label{3.12}
\end{equation}

At this point it is convenient to split the Lagrangian (\ref{3.10}) in two independent parts according to
\begin{eqnarray}
\mathcal{L}_{gapless}&\equiv&\sum_{\sigma,\rho=1}^2\sum_{i,j=1}^k \left[\psi_{R,\sigma,i}^{1\ast}
[(\mathrm{i}\partial_+ +A_+^1)\delta_{\sigma\rho}\delta_{ij}+(B_+^1)^{ij}\delta_{\sigma\rho} +(C_+^1)^{\sigma\rho}\delta_{ij}]
\psi_{R,\rho,j}^1 \right.\nonumber\\
&+&\left.\psi_{L,\sigma,i}^{\mathcal{N}\ast}[(\mathrm{i}\partial_{-}+A_-^{\mathcal{N}})\delta_{\sigma\rho}\delta_{ij}+
(B_-^{\mathcal{N}})^{ij}\delta_{\sigma\rho}
+(C_{-}^{\mathcal{N}})^{\sigma\rho}\delta_{ij}]\psi_{L,\rho,j}^{\mathcal{N}}\right]\nonumber\\
&+& \sum_{\sigma\rho=1}^2\sum_{i',j'=1}^{k'}\left[\psi_{R,\sigma,i'}^{1\ast}[(\mathrm{i}\partial_{+}+A^{\prime 1}_+)\delta_{\sigma\rho}\delta_{i'j'}
+(B_+^{\prime 1})^{ i'j'}\delta_{\sigma\rho}+(C_+^1)^{\sigma\rho}\delta_{i'j'}]\psi_{R,\rho,j'}^1 \right.\nonumber\\
&+&\left.\psi_{L,\sigma,i'}^{\mathcal{N}\ast}[(\mathrm{i}\partial_{-}+A^{\prime \mathcal{N}}_-)\delta_{\sigma\rho}\delta_{i'j'}+
(B_-^{\prime \mathcal{N}})^{i'j'}\delta_{\sigma\rho}+(C_-^{\mathcal{N}})^{\sigma\rho}\delta_{i'j'}]\psi_{L,\rho,j'}^{\mathcal{N}}\right]
\label{3.10a}
\end{eqnarray}
and
\begin{eqnarray}
\mathcal{L}_{gapped}&\equiv&\sum_{\sigma,\rho=1}^2\sum_{i,j=1}^k \left[\psi_{R,\sigma,i}^{\mathcal{N}\ast}
[(\mathrm{i}\partial_+ +A_+^{\mathcal{N}})\delta_{\sigma\rho}\delta_{ij}+(B_+^{\mathcal{N}})^{ij}\delta_{\sigma\rho} +(C_+^{\mathcal{N}})^{\sigma\rho}\delta_{ij}]
\psi_{R,\rho,j}^{\mathcal{N}} \right.\nonumber\\
&+&\left.\psi_{L,\sigma,i}^{1\ast}[(\mathrm{i}\partial_{-}+A_-^1)\delta_{\sigma\rho}\delta_{ij}+
(B_-^1)^{ij}\delta_{\sigma\rho}
+(C_{-}^1)^{\sigma\rho}\delta_{ij}]\psi_{L,\rho,j}^1\right]\nonumber\\
&+& \sum_{\sigma\rho=1}^2\sum_{i',j'=1}^{k'}\left[\psi_{R,\sigma,i'}^{\mathcal{N}\ast}[(\mathrm{i}\partial_{+}+A^{\prime \mathcal{N}}_+)\delta_{\sigma\rho}\delta_{i'j'}
+(B_+^{\prime \mathcal{N}})^{ i'j'}\delta_{\sigma\rho}+(C_+^{\prime \mathcal{N}})^{\sigma\rho}\delta_{i'j'}]\psi_{R,\rho,j'}^{\mathcal{N}} \right.
\nonumber\\
&+&\left.\psi_{L,\sigma,i'}^{1\ast}[(\mathrm{i}\partial_{-}+A^{\prime 1}_-)\delta_{\sigma\rho}\delta_{i'j'}+
(B_-^{\prime 1})^{i'j'}\delta_{\sigma\rho}+(C_-^{\prime 1})^{\sigma\rho}\delta_{i'j'}]\psi_{L,\rho,j'}^1\right].
\label{3.10b}
\end{eqnarray}
As the subscripts already indicate, these contributions are divided  
according to the presence or not of gapless degrees of freedom. 
Thus our next task is to show that the Lagrangian (\ref{3.10a}) contains the gapless degrees of freedom whereas 
the Lagrangian (\ref{3.10b}) is fully gapped. 

%%%%%%%%%%%%%%%%%%%%%%%%%%%%%%%%%%%%%%%%%%%%%%%%%%%%%%%%%%
\subsection{Gapless Sector}

In order to extract the gapless physical content of (\ref{3.10a}) we have to deal with the gauge redundancy that it contains.
At the first sight it is invariant under independent transformations for right and left fields. For example, for the
fermions of the $k$-wires transforming as $\psi_{R}^{1}\rightarrow e^{\mathrm{i} \Lambda_{R}}\psi_{R}^{1}$ and 
$\psi_{L}^{\mathcal{N}}\rightarrow e^{\mathrm{i} \Lambda_{L}}\psi_{L}^{\mathcal{N}}$
and the Abelian fields as $A_{+}^1\rightarrow A_{+}^1+ \partial_{+} \Lambda_{R}$ and $A_{+}^{\mathcal{N}}\rightarrow A_{+}^{\mathcal{N}}+ \partial_{+} \Lambda_{L}$, with
different functions $\Lambda_{R}$ and $\Lambda_{L}$.     
However, such invariance is an illusion since it fails at the quantum level --- the integration measure 
is not invariant and the theory contains an anomaly. 
Rather it is invariant under transformations with the same function for right and left sectors $\Lambda_{R}=\Lambda_{L}\equiv \Lambda$, 
once the anomalies in both sectors are canceled. In this way, the gauge 
invariance ties the fields of first and last bundle. This is
how the existence of the bulk manifest itself into the edges.  

To avoid unnecessary heavy notation we omit the index bundle of the fields without 
prejudice of ambiguities.
The full set of gauge transformations which leave invariant the partition function associated to the Lagrangian (\ref{3.10a}) are: 
\begin{equation}
\psi_{R/L,\sigma,i} \rightarrow e^{\mathrm{i} \Lambda}\psi_{R/L,\sigma,i} ~~~\text{and}~~~ A_{\pm}\rightarrow A_{\pm}+ \partial_{\pm} \Lambda;
\label{3.13}
\end{equation}
\begin{equation}
\psi_{R/L,\sigma,i'} \rightarrow e^{\mathrm{i} \Lambda'}\psi_{R/L,\sigma,i'} ~~~\text{and}~~~ A'_{\pm}\rightarrow A'_{\pm}+ \partial_{\pm} \Lambda';
\label{3.14}
\end{equation} 
\begin{equation}
\psi_{R/L,\sigma,i} \rightarrow M_{ij}\psi_{R/L,\sigma,j},~~~ B_{\pm}\rightarrow M B_{\pm} M^{-1} - \partial_{\pm}M M^{-1}, ~~~M\in SU(k);
\label{3.15}
\end{equation}
\begin{equation}
\psi_{R/L,\sigma,i'} \rightarrow M'_{i'j'}\psi_{R/L,\sigma,j'},~~~ B'_{\pm}\rightarrow M' B'_{\pm} M^{\prime-1} - \partial_{\pm}M'; 
\label{3.16}
M^{\prime-1}, ~~~M'\in SU(k');
\end{equation}
\begin{equation}
\psi_{R/L,\sigma,i} \rightarrow N_{\sigma\rho}\psi_{R/L,\rho,i},~~\psi_{R/L,\sigma,i'} \rightarrow N_{\sigma\rho}\psi_{R/L,\rho,i'},
~~ C_{\pm}\rightarrow N C_{\pm} N^{-1} - \partial_{\pm}N N^{-1}, ~~N\in SU(2).
\label{3.17}
\end{equation} 
Note that we do not have independent $SU(2)$ $C$-fields for wires belonging to sets $k$ and $k'$. As we shall see now, 
this is what allows for the existence of gapless degrees of freedom.
In order to have a well-defined partition function 
we need to fix the gauge in the path integral (\ref{3.9}) according to the Faddeev-Popov procedure. This involves the introduction
of ghost fields, which are responsible for subtracting central charge since they effectively counts as a
negative number of degrees of freedom. We proceed by fixing the gauge as 
\begin{equation}
A_-=0~~~\text{and}~~~A'_-=0;
\label{3.18}
\end{equation}
\begin{equation}
B_-=0~~~\text{and}~~~B'_-=0;
\label{3.19}
\end{equation}
\begin{equation}
C_-=0.
\label{3.20}
\end{equation}
The corresponding Faddeev-Popov determinants can be exponentiated in terms of $b$-$c$ ghost systems:
\begin{equation}
\mathcal{L}_{FP}= \underbrace{b\partial_{-} c}_{A_-=0} + \underbrace{b'\partial_{-} c'}_{A'_-=0} +
\underbrace{\sum_{i=1}^{k^2-1} b_i\partial_{-} c_i}_{B_-=0}+\underbrace{\sum_{i'=1}^{k^{\prime 2}-1} b'_{i'}\partial_{-} c'_{i'}}_{B'_-=0}+
\underbrace{\sum_{\sigma=1}^{3}b_{\sigma}\partial_{-} c_{\sigma}}_{C_-=0},
\label{3.21}
\end{equation}
where the ghost fields $b$'s and $c$'s have conformal weights $h_b=1$ and $h_c=0$, respectively. Each ghost field 
gives a central charge $-2$ for the right sector, such that the total contribution of the ghost action is
\begin{equation}
-2-2- (k^2-1)\times 2 -(k^{\prime 2}-1)\times 2 - 3\times 2 =-2(k^2+k^{\prime 2}) -6.
\label{3.21a}
\end{equation}
Note that the ghost action does not give any contribution for the central charge of the left sector. 
This imbalance of central charge in the two sectors is equalized  
with the contributions discussed below. 
The remaining gauge fields in the Lagrangian (\ref{3.10}) can be parametrized as  
\begin{equation}
A_+ = \mathrm{i} \phi^{-1}\partial_{+} \phi ~~~\text{and}~~~A'_+ = \mathrm{i} \phi^{\prime-1}\partial_{+} \phi',~~~ \phi, \phi'\in U(1);
\label{3.22}
\end{equation}
\begin{equation}
B_+ = \mathrm{i} h^{-1}\partial_{+} h,~~~h\in SU(k) ~~~\text{and}~~~B'_+ = \mathrm{i} h^{\prime-1}\partial_{+} h',~~~h'\in SU(k');
\label{3.23}
\end{equation}
\begin{equation}
C_+ = \mathrm{i} g^{-1}\partial_{+} g,~~~ g\in SU(2).
\label{3.24}
\end{equation}
These changes of variables produce nontrivial Jacobians which will be taken into account in a moment. 
As discussed in the Appendix \ref{A1}, 
the integration over fermionic fields produces
\begin{eqnarray}
&&\int \mathcal{D}\psi_R \mathcal{D}\psi_L \exp \mathrm{i} \int dt dx \,\mathcal{\tilde{L}}[\psi_R,\psi_L, A_+,A'_+,B_+,B'_+,C_+]\nonumber\\
&=& \underbrace{\det\partial_+\det\partial_-}_{k~ \text{times}}\underbrace{\det\partial_+\det\partial_-}_{k'~\text{times}}
\exp  \mathrm{i}\left(- S[\phi] - S[\phi'] -  2 S[h] - 2S[h'] - (k+k') S[g] \right),
\label{3.25}
\end{eqnarray}
where $S[h]$ is the WZW action
\begin{equation}
S[h]=\frac{1}{8\pi} \int d^2x \text{Tr} \,\partial_{\mu }h\partial^{\mu} h^{-1} +\frac{1}{12\pi} \int d^3x 
 \epsilon^{\mu\nu\rho} \text{Tr}\,h^{-1}\partial_{\mu} h\, h^{-1}\partial_{\nu} h\, h^{-1}\partial_{\rho} h.
 \label{3.26}
\end{equation}
It defines a two dimensional quantum field theory in spite of the presence of a three-dimensional integral. Indeed, this term is a Wess-Zumino term that encodes the chiral anomaly in two-dimensional spacetime \cite{Nair}.
For the Abelian gauge fields it reduces to the Lagrangian of a free scalar field, which can be easily seen if 
we write $\phi = e^{\mathrm{i} \varphi}$. The determinants $\det \partial_{+}$ and $\det \partial_{-} $ in (\ref{3.25}) can be exponentiated back
in terms of the functional integration over the original fermionic fields $\psi_{R/L}$.

The Jacobians for the Abelian parametrizations in (\ref{3.22}) are
\begin{equation}
\mathcal{D}A_+ = (\det\partial_{+})_{U(1)}\mathcal{D}\phi~~~\text{and}~~~\mathcal{D}A'_+ = (\det\partial_{+})_{U(1)}\mathcal{D}\phi',
\label{3.27a}
\end{equation} 
while the non-Abelian transformations (\ref{3.23}) and (\ref{3.24}) possess nontrivial contributions
\begin{equation}
\mathcal{D}B_+ = (\det\partial_{+})_{SU(k)}\,e^{ - \mathrm{i} 2k S[h]}\mathcal{D}h, 
~~~\mathcal{D}B'_+ =(\det\partial_{+})_{SU(k')} \,e^{ - \mathrm{i} 2 k'  S[h']}\mathcal{D}h',
\label{3.27b}
\end{equation}
and
\begin{equation}
\mathcal{D}C_+ =(\det\partial_{+})_{SU(2)}\, e^{ - \mathrm{i} 4 S[g]}\mathcal{D}g.
\label{3.27c}
\end{equation}
These determinants can be exponentiated in terms of ghost fields that subtract central charge of the left sectors, 
restoring the equality of central charge in the two sectors. 
Collecting all the contributions we obtain the final form for partition function  
\begin{equation}
\mathcal{Z}_{gapless}= \int \mathcal{D}\psi_R \mathcal{D}\psi_L \mathcal{D}\phi \mathcal{D}\phi' \mathcal{D}h \mathcal{D}h'\mathcal{D}g
\mathcal{D}\mathcal{G} \exp \mathrm{i} S_{eff}, 
\label{3.28}
\end{equation}
with the measure $\mathcal{D}\mathcal{G}$ representing the functional integral over all ghost fields 
in the action (\ref{3.21}) --- right and left sectors --- and the effective action is given by 
\begin{equation}
S_{eff}= S[\psi_R,\psi_L]-S[\phi]-S[\phi']- (2+2k) S[h]- (2+2k') S[h']-(k+k'+4)S[g]+S_{ghost}.
\label{3.29}
\end{equation}
The first term
$S[\psi_R,\psi_L]$ is the action associated with the Lagrangian (\ref{3.10a})  and $S_{ghost}$ includes the Lagrangian (\ref{3.21})
and the corresponding left sector ghost fields.
From the effective action we can determine the total central charge by adding the contributions of the parts, remembering that 
the central charge for the WZW theory is given by the Knizhnik-Zamolodchikov formula \cite{Knizhnik}
\begin{equation}
c_{WZW}= \frac{p \,\text{dim}H}{p+C_H},
\label{3.29a}
\end{equation}
where $p$ is the Kac-Moody level of the WZW currents and $C_H$ is the quadratic Casimir of the algebra $H$ --- for $SU(N)$ the quadratic Casimir is $C_H=N$.
The resulting central charge is
\begin{eqnarray}
c_{R/L}(k,k')&=&c_{matter}+c_{WZW}+c_{ghost}\nonumber\\
&=&2 (k + k') + 1 + 1 + \frac{(2 + 2 k) (k^2 - 1)}{(
  2 + k)} +\frac{ (2 + 2 k') (k'^2 - 1)}{(2 + k')} + \frac{(k + k' + 4) 3}{(k + k' + 2)} \nonumber\\&-& 
 2 (k^2 + k'^2) - 6.
 \label{3.30}
\end{eqnarray}
It is easy to see that this central charge is always smaller than the central charge of the free fermions inside the bundle
$2(k+k')$ and is equal to the one in expression (\ref{2.23}). We  see that the effect of constraining currents is to remove gapless degrees of freedom 
since they introduce some gauge redundancy. The constraints on the currents effectively implement
the interactions between nearby wires.

We shall discuss now the issue of chiral splitting of the action (\ref{3.29}). 
This is important since the global picture of the topological phase corresponds to each chiral 
conformal field theory living at one of the edges.  
In decomposing the action  (\ref{3.29}) into two chiralities we need to deal with four type of theories. 
Decomposition of fermions, Abelian fields and ghosts are straightforward. 
Thus we will focus on the WZW contribution (\ref{3.26}).
 
 At first we consider the nonlinear sigma model part of WZW action and write it in terms of 
an auxiliary field $\Lambda$ \cite{Braga},
\begin{equation}
S_{NSM}= \frac{1}{4\pi}\int d^2x \text{Tr}\left( \frac{1}{2}\Lambda h \Lambda h +\partial_t h \Lambda+
\frac{1}{2}h^{-1}\partial_x h h^{-1}\partial_x h \right),
\label{c1.1}
\end{equation}
where we have taken our two dimensional Minkowski metric as $(1,-1)$. By using the equation
of motion for $\Lambda$ and plugging it back into the action we restore the original form of the nonlinear sigma model.
Next we introduce new fields $M_R$ and $M_L$ as
\begin{equation}
h\equiv M_L M_R~~~\text{and}~~~ \Lambda\equiv -M_R^{-1}\partial_x M_L^{-1}+\partial_x M_R^{-1} M_L^{-1}.
\label{c1.2}
\end{equation}   
It is straightforward to show that the action becomes
\begin{eqnarray}
S_{NSM}&=&-\frac{1}{4\pi}\int d^2x \text{Tr}\left[ \partial_x M_L^{-1} (\partial_t+\partial_x)M_L\right] +
\frac{1}{4\pi}\int d^2x \text{Tr}\left[ \partial_x M_R^{-1} (\partial_t-\partial_x)M_R\right]\nonumber\\
&+&\frac{1}{4\pi} \int d^2x \text{Tr} \left[M_L^{-1}(\partial_t M_L \partial_x M_R- \partial_x M_L \partial_t M_R)M_R^{-1}\right].
\label{c1.3}
\end{eqnarray}
Note that at this point there is no yet a chiral splitting due to the mixing term in the last line. It is exactly canceled when 
we include the topological term
\begin{equation}
\Gamma[h]\equiv \frac{1}{12\pi} \int d^3x 
 \epsilon^{\mu\nu\rho} \text{Tr}\,h^{-1}\partial_{\mu} h\, h^{-1}\partial_{\nu} h\, h^{-1}\partial_{\rho} h.
 \label{c1.4}
\end{equation}
Indeed, by setting $h= M_L M_R$ in (\ref{c1.4}) it follows that,
\begin{equation}
\Gamma[M_L M_R]=\Gamma[M_L]+\Gamma[M_R]-
\frac{1}{4\pi} \int d^2x \text{Tr} \left[M_L^{-1}(\partial_t M_L \partial_x M_R- \partial_x M_L \partial_t M_R)M_R^{-1}\right].
\label{c1.5}
\end{equation}
By considering the full WZW action, $S=S_{NSM}+\Gamma$, we obtain 
\begin{eqnarray}
S&=& -\frac{1}{4\pi}\int d^2x \text{Tr}\left[ \partial_x M_L^{-1} (\partial_t+\partial_x)M_L\right]+\Gamma[M_L]\nonumber\\
&+&\frac{1}{4\pi}\int d^2x \text{Tr}\left[ \partial_x M_R^{-1} (\partial_t-\partial_x)M_R\right] +\Gamma[M_R],
\label{c1.6}
\end{eqnarray}
which is the chiral-split WZW action desired. Thus we see that the chiral gapless degrees of freedom characterized by the 
central charges $c_{R}$ and $c_{L}$ of equation (\ref{3.30}) are decoupled and propagate 
at the opposite edges of the system.

%As discussed previously, when we are considering the 1+1 dimensional theory itself, we keep both chiralities in coupling fermions with gauge fields to avoid issues with gauge anomalies. On the other hand, viewed as the boundary of a 2+1 theory, we can have only one of the chiralities in (\ref{c1.6}), since the anomaly of the edge is canceled against the anomaly of the bulk. This is the basis of bulk-edge correspondence  and will be discuss later. 

%%%%%%%%%%%%%%%%%%%%%%%%%%%%%%%%%%%%%%%
\subsection{Gapped Sector}

To show that the Lagrangian (\ref{3.10b}) is fully gapped we follow the  same general strategy  
of the previous section but with the crucial difference that now the involved
$SU(2)$ $C$-fields of the sets of $k$ and $k'$ wires are independent.
This implies that we have two independent gauge fixing conditions
\begin{equation}
C_{-}=0 ~~~\text{and}~~~ C'_{-}=0
\label{g1}
\end{equation} 
and correspondingly ghosts fields for both of them, and also two 
independent parametrizations  
\begin{equation}
C_{+}=\mathrm{i} g^{-1}\partial_{+} g,~~~\text{and}~~~ C'_{+}=\mathrm{i} g^{\prime -1}\partial_{+} g',~~~
g,g'\in SU(2).
\label{g2}
\end{equation}
By following the same steps as before, we arrive at
\begin{equation}
\mathcal{Z}_{gapped}= \int \mathcal{D}\psi_R \mathcal{D}\psi_L \mathcal{D}\phi \mathcal{D}\phi' \mathcal{D}h \mathcal{D}h'\mathcal{D}g \mathcal{D}g'
\mathcal{D}\mathcal{G} \exp \mathrm{i} S_{eff}, 
\label{g3}
\end{equation}
where the effective action is given now by 
\begin{equation}
S_{eff}= S[\psi_R,\psi_L]-S[\phi]-S[\phi']- (2+2k) S[h]- (2+2k') S[h']-(k+4)S[g]-(k'+4)S[g']+S_{ghost}.
\label{g4}
\end{equation}
It is straightforward to show that the resulting central charge vanishes,
\begin{eqnarray}
c_{R/L}(k,k')&=&c_{matter}+c_{WZW}+c_{ghots}\nonumber\\
&=& 2 (k + k') + 1 + 1 + \frac{(2 + 2 k) (k^2 - 1)}{(
  2 + k)} + \frac{(2 + 2 k') (k^{\prime 2} - 1)}{(2 + k')} + \frac{(k + 4) 3}{(
  k + 2)} + \frac{(k' + 4) 3}{(k' + 2)}\nonumber\\ 
  &-& 2 (k^2 + k^{\prime 2}) - 6 - 6\nonumber\\
 &=&0.
\label{g5}
\end{eqnarray}
Thus, as anticipated in the previous section, there is no gapless degrees of freedom associated to the Lagrangian 
(\ref{3.10b}). All contributions for the conformal sector are due to the Lagrangian (\ref{3.10a}). 

To sum up, these results show that the effective action (\ref{3.29}) coming from fermions with constrained currents
represents the edge states obtained in the coupled wires approach of Sec. \ref{Sec1}. However, it is still not a convenient form to make connection with 
the topological bulk theory since it is hard to obtain from a bulk theory a boundary theory like (\ref{3.29}) involving 
diverse types of fields. 
We need one more step to show that it is equivalent to a gauged version 
of chiral WZW model, which will be discussed in next section.

%%%%%%%%%%%%%%%%%%%%%%%%%%%%%%%%%%%%%%%%%%%
\section{Gauged WZW Models}\label{CGWZ}

We will discuss now in some detail the equivalence between the gapless sector of the constrained fermionic theory of the previous section and
a gauged WZW model following Ref. \cite{Cabra}.  We start by considering the equivalence for the case where the gauge fields are background fields, i.e., 
they are not integrated in the 
partition function,
\begin{equation}
Z[A_{\pm},A_{\pm}',B_{\pm},B_{\pm}',C_{\pm}]=\int \mathcal{D}\psi_R \mathcal{D}\psi_L \exp \mathrm{i} \tilde{S}[\psi_R,\psi_L,A_{\pm},A_{\pm}',B_{\pm},B_{\pm}',C_{\pm}],
\label{3.34}
\end{equation}
with $\tilde{S}$ being the action corresponding to the Lagrangian (\ref{3.10a}). Let us assume that the partition function is normalized 
to the unity in absence of background fields.
Next, instead of fixing the gauge by putting the ($-$) components of the gauge fields equal to zero, we parametrize them similarly to 
(\ref{3.22}), (\ref{3.23}) and (\ref{3.24}):
\begin{equation}
A_- = \mathrm{i} \bar{\phi}^{-1}\partial_{-} \bar{\phi} ~~~\text{and}~~~A'_- = \mathrm{i} \bar{\phi}^{\prime-1}\partial_{-} \bar{\phi}',~~~ \bar{\phi}, \bar{\phi}'\in U(1);
\label{3.35}
\end{equation}
\begin{equation}
B_- = \mathrm{i} \bar{h}^{-1}\partial_{-} \bar{h},~~~\bar{h}\in SU(k) ~~~\text{and}~~~B'_- = \mathrm{i} \bar{h}^{\prime-1}\partial_{-} \bar{h}',~~~\bar{h}'\in SU(k');
\label{3.36}
\end{equation}
\begin{equation}
C_- = \mathrm{i} \bar{g}^{-1}\partial_{-} \bar{g},~~~ \bar{g}\in SU(2).
\label{3.37}
\end{equation}
Now let us define the fields 
\begin{equation}
U\equiv \phi\otimes h\otimes g~~~\text{and}~~~U'\equiv \phi'\otimes h'\otimes g
\label{3.38}
\end{equation}
and 
\begin{equation}
\bar{U}\equiv \bar{\phi}\otimes \bar{h}\otimes \bar{g}~~~\text{and}~~~\bar{U}'\equiv \bar{\phi}'\otimes \bar{h}'\otimes \bar{g}.
\label{3.39}
\end{equation}
The partition function can be written as 
\begin{eqnarray}
&&Z[A_{\pm},A_{\pm}',B_{\pm},B_{\pm}',C_{\pm}]= \int \mathcal{D}\psi_R \mathcal{D}\psi_L \exp \mathrm{i}\int dt dx \nonumber\\ 
&\times&\left\{ \sum_{\sigma,\rho=1}^2\sum_{i,j=1}^k \left[\psi_{R,\sigma,i}^{\ast}
[(\mathrm{i}\partial_{+} + \mathrm{i}U^{-1}\partial_{+}U)_{ij,\sigma\rho}]
\psi_{R,\rho,j} +
\psi_{L,\sigma,i}^{\ast}[(\mathrm{i}\partial_{-}+ \mathrm{i} \bar{U}^{-1}\partial_{-}\bar{U})_{ij,\sigma\rho}]\psi_{L,\rho,j}\right]\right.\nonumber\\
&+& \left.\sum_{\sigma\rho=1}^2\sum_{i',j'=1}^{k'}\left[\psi_{R,\sigma,i'}^{\ast}
[(\mathrm{i}\partial_{+}+\mathrm{i}U^{\prime -1}\partial_{+}U')_{i'j',\sigma\rho}]\psi_{R,\rho,j'} +\psi_{L,\sigma,i'}^{\ast}[(\mathrm{i}\partial_{-}+\mathrm{i}\bar{U}'\partial_{-}\bar{U}')_{i'j',\sigma\rho}]\psi_{L,\rho,j'}\right]\right\}.\nonumber\\
\label{3.40}
\end{eqnarray}
Note that we can introduce fields as Kronecker sums\footnote{Our notation for Kronecker sum is $A\oplus B\equiv A\otimes I + I\otimes B$. For  
direct sum of matrices we write with big symbol $A\bigoplus B$.}
\begin{equation}
\mathcal{A}_{\pm}\equiv A_{\pm}\oplus  B_{\pm}\oplus C_{\pm}~~~\text{and}~~~
\mathcal{A}'_{\pm}\equiv A'_{\pm}\oplus B'_{\pm}\oplus C_{\pm},
\label{3.40a}
\end{equation}
such that 
\begin{equation}
\mathcal{A}_{+}=\mathrm{i} U^{-1}\partial_{+}U~~~\text{and}~~~\mathcal{A}_{-}=\mathrm{i} \bar{U}^{-1}\partial_{+}\bar{U}
\label{3.40b}
\end{equation}
and
\begin{equation}
\mathcal{A}'_{+}=\mathrm{i} U^{\prime-1}\partial_{+}U'~~~\text{and}~~~\mathcal{A}'_{-}=\mathrm{i} \bar{U}^{\prime-1}\partial_{+}\bar{U}'.
\label{3.40c}
\end{equation}
Integration over fermions produces
\begin{eqnarray}
Z[\mathcal{A}_{\pm},\mathcal{A}'_{\pm}]&=& \exp\left[ -\mathrm{i} S[U \bar{U}^{-1}]+\mathrm{i} \alpha \int dt dx \text{Tr} \mathcal{A}_{+}
\mathcal{A}_{-}\right]
 \nonumber\\
&\times& \exp\left[ -\mathrm{i} S[U' \bar{U}^{\prime-1}]+\mathrm{i} \alpha' \int dt dx \text{Tr} \mathcal{A}'_{+}
\mathcal{A}'_{-}\right],
\label{3.41}
\end{eqnarray}
with $\alpha$ and $\alpha'$ being regularization-dependent parameters \cite{Jackiw}. 
This expression can be compared with that one coming from the gauged-WZW model, 
defined by the action
\begin{eqnarray}
S_{gauge}[w,F_{\pm}]\equiv S[w]+\frac{1}{4\pi}\int dt dx \text{Tr} \left[
F_{+}\partial_{-}w w^{-1}-F_{-}w^{-1}\partial_{+}w
+ F_{+} w F_{-}w^{-1}-F_{+}F_{-}\right],
\label{3.41a}
\end{eqnarray}
where $w$ is a matrix-valued field belonging to some group $G$ and $F_{\pm}$ take values in a subgroup $H$ of $G$. 
The gauged-WZW model that we consider is written in terms of the direct sum  $w\bigoplus w'$, with $w\in U(2k)$ and $w'\in U(2k')$, coupled to the direct sum of the background fields  (\ref{3.40a}), 
$\mathcal{A}_{\pm}\bigoplus \mathcal{A}'_{\pm}$. The action becomes
\begin{equation}
S_{gauge}[w\bigoplus w',\mathcal{A}_{\pm}\bigoplus \mathcal{A}'_{\pm}]=S_{gauge}[w,\mathcal{A}_{\pm}]+
S_{gauge}[w',\mathcal{A}'_{\pm}].
\label{3.42}
\end{equation}
The partition function in the presence of background fields is given by
\begin{equation}
\tilde{Z}[\mathcal{A}_{\pm},\mathcal{A}'_{\pm}]=\int \mathcal{D}w \mathcal{D}w' 
e^{\mathrm{i} S_{gauge}[w,\mathcal{A}_{\pm}]}e^{i S_{gauge}[w',\mathcal{A}'_{\pm}]},
\label{3.43}
\end{equation}
which is assumed to be normalized to the unity in the absence of background fields. 
Now we can use the Polyakov-Wiegmann identity \cite{Polyakov},
\begin{equation}
S[MN]=S[M]+S[N]-\frac{1}{4\pi}\int{d^2x\text{tr}\left(M^{-1}\partial_{+}M\partial_{-}NN^{-1}\right)},
\label{3.44}
\end{equation}
to obtain
\begin{equation}
S_{gauge}[w,\mathcal{A}_{\pm}]=S[U w \bar{U}^{-1}]-S[U \bar{U}^{-1}].
\label{3.45}
\end{equation}
With this the partition function (\ref{3.42}) becomes
\begin{equation}
\tilde{Z}[\mathcal{A}_{\pm},\mathcal{A}'_{\pm}] =\int \mathcal{D}w \mathcal{D}w' e^{\mathrm{i} S[U w \bar{U}^{-1}]-\mathrm{i} S[U \bar{U}^{-1}]}
e^{\mathrm{i} S[U' w' \bar{U}^{\prime-1}]-\mathrm{i} S[U' \bar{U}^{\prime-1}]}.
\label{3.46}
\end{equation}
With the change of variables $U w \bar{U}^{-1}\rightarrow w$ and $U' w' \bar{U}^{\prime-1}\rightarrow w'$
we see that the partition function factorizes and as it is normalized we get
\begin{equation}
\tilde{Z}[\mathcal{A}_{\pm},\mathcal{A}'_{\pm}]=e^{-\mathrm{i} S[U \bar{U}^{-1}]} e^{-\mathrm{i} S[U' \bar{U}^{\prime-1}]}.
\label{3.47}
\end{equation}
Thus if the regularization is chosen such that $\alpha=\alpha'=0$ in (\ref{3.41}) we obtain the 
equivalence between the two models. 

The equivalence to the case of dynamical fields is immediate.  
Just replace the fields $\mathcal{A}_{\pm}$ and $\mathcal{A}'_{\pm}$ in the fermionic partition function (\ref{3.40})
by $E_{\pm}+\mathcal{A}_{\pm}$ and $E'_{\pm}+\mathcal{A}'_{\pm}$ where the fields $E\in U(2k)$ and $E'\in U(2k')$
and then integrate over the gauge fields. Integration over the fermions furnishes
\begin{equation}
\mathcal{Z}[E_{\pm},E'_{\pm}]\equiv \int \mathcal{D}\mathcal{A}_{\pm} \mathcal{D}\mathcal{A}'_{\pm}
Z[E_{\pm}+\mathcal{A}_{\pm},E'_{\pm}+\mathcal{A}'_{\pm}].
\label{3.48}
\end{equation}

By proceeding in a similar way for the partition function (\ref{3.42}), it follows that
\begin{equation}
\tilde{\mathcal{Z}}[E_{\pm},E'_{\pm}]\equiv \int \mathcal{D}\mathcal{A}_{\pm} \mathcal{D}\mathcal{A}'_{\pm}
\tilde{Z}[\mathcal{A}_{\pm}+E_{\pm},\mathcal{A}'_{\pm}+E'_{\pm}].
\label{3.49}
\end{equation}
Since we have shown that the integrand of the two above partition functions are the same, we conclude that 
$\mathcal{Z}[E_{\pm},E'_{\pm}]=\tilde{\mathcal{Z}}[E_{\pm},E'_{\pm}]$. In other words, 
the fermionic theory with constraints over the currents (\ref{3.9}) is equivalent to the gauged-WZW theory 
\begin{equation}
\tilde{\mathcal{Z}}=\int \mathcal{D}w \mathcal{D}w' \mathcal{D}\mathcal{A}_{\pm} \mathcal{D}\mathcal{A}'_{\pm}
e^{\mathrm{i} S_{gauge}[w,\mathcal{A}_{\pm}]}e^{\mathrm{i} S_{gauge}[w',\mathcal{A}'_{\pm}]},
\label{3.50}
\end{equation}
in the sense that both of them produce the same correlation functions. 
Our next task is to show that this gauged-WZW theory follows naturally from a coset Chern-Simons-like bulk theory 
whenever we define it in the presence of a physical edge.

%%%%%%%%%%%%%%%%%%%%%%%%%%%%%%%%%%%%%%%%%%%%%
%%%%%%%%%%%%%%%%%%%%%%%%%%%%%%%%%%%%%%%%%%%%
\section{Effective Field Theory}\label{Sec3}

In Secs. \ref{Sec1} and \ref{Sec2}, we have derived the properties of the gapped topological phases starting from a fermionic interacting quantum field theory. In this section we will focus on a more effective approach that exhibits the important properties of the ground states of the gapped phases, namely only the topological ones. It turns out that the Chern-Simons theories captures the relevant topological aspects of the low-energy phenomenology of these systems. 

In particular, a common feature of these topological phases is the appearance of gapless excitations at the boundary of the manifold where the system is placed. The relation between the bulk dynamics with the boundary dynamics can be obtained through the  bulk-edge correspondence \cite{Wen2,Wen3}. Both, the Abelian and non-Abelian Chern-Simons theories, exhibit such a correspondence and are important in the description of the aspects of the topological phases of matter. However, since our interest is on the phases described by a coset of non-Abelian algebras, we will focus only 
on the correspondence between non-Abelian Chern-Simons theories and the boundary WZW models and their gauged versions for the case of coset of non-Abelian algebras.

%%%%%%%%%%%%%%%%%%%%%%%%%%%%%%%%%%%%%%%%%%%%%%%%%%%%
\subsection{Non-Abelian Chern-Simons}

To describe conformal field theories with non-integer values for the central charge on the boundary in terms of a bulk topological field theory we need to consider non-Abelian Chern-Simons models.
We consider a non-Abelian unitary group $G$ with generators $G^\dagger_a=-G_a$. 
For a Lie-algebra-valued gauge field $A_\mu\equiv A^a_\mu G_a$ the non-Abelian Chern-Simons classical model is defined by the action
\begin{subequations}
\begin{equation}
S_{CS}[A]=\frac{k}{4\pi}\int{\epsilon^{\mu\nu\rho}\text{Tr}\left[A_\mu\partial_\nu A_\rho+\frac{2}{3}A_\mu A_\nu A_\rho\right]},\label{non ab cs action}
\end{equation}
with the group generators satisfying the algebra
\begin{equation}
[G_a,G_b]=f_{abc}G_c.
\end{equation}
We adjust the generator basis to have completely antisymmetric structure constants $f_{abc}$ and to obey
\begin{equation}
\text{Tr}\left(G_aG_b\right)=-\frac{1}{2}\delta_{ab}.
\end{equation}
\end{subequations}

Under a gauge transformation,
\begin{equation}
A_{\mu}\rightarrow A^{\prime}_\mu=U^{-1}A_\mu U+ U^{-1}\partial_\mu U,
\label{non ab gauge}
\end{equation}
the action (\ref{non ab cs action}) changes to
\begin{equation}
S_{CS}\rightarrow S_{CS}-\frac{k}{12\pi}\int_{\Omega}{\epsilon^{\mu\nu\rho}\text{Tr}\left(U^{-1}\partial_\mu UU^{-1}\partial_\nu UU^{-1}\partial_\rho U\right)}+ \frac{k}{4\pi}\int_{\Omega}{\epsilon^{\mu\nu\rho}\partial_\nu\left(\partial_\mu UU^{-1}A_\rho\right)}.
\label{non ab cs action gauge}
\end{equation}
For trivial boundary conditions the last integral vanishes by using the Gauss theorem. We can also consider the compactification of the $2+1$ 
spacetime to the $3$-sphere $S^3$ by demanding
\begin{equation}
U(x)\xrightarrow{\quad\partial \Omega\quad} I.
\label{identity}
\end{equation}
In this case, the group elements $U(x)$ are maps from $S^3$ to $G$. These maps are topologically classified by the third homotopy group $\Pi_3(G)$. With exception of $SO(4)$, which has $\Pi_3(SO(4))=\mathbb{Z}\times\mathbb{Z}$, all other simple non-Abelian groups have $\Pi_3(G)=\mathbb{Z}$. The first integral in (\ref{non ab cs action gauge}) can be recognized as the winding number for such maps,
\begin{equation}
w(U)=-\frac{1}{24\pi^2}\oint_{S^3}{\epsilon^{\mu\nu\rho}\text{Tr}\left(U^{-1}\partial_\mu UU^{-1}\partial_\nu UU^{-1}\partial_\rho U\right)},\label{w number}
\end{equation}
which counts the number of times the map winds a target $3$-sphere, associated to a $SU(2)$ subgroup of $G$, while the domain $3$-sphere is covered once.

For the group elements that are continuously connected to the identity, the winding number is zero, and the action (\ref{non ab cs action}) is then invariant under such a subset of transformations. However, for the more general transformations called large gauge transformations, the winding number is a non-zero integer. Therefore, the action is not gauge invariant for these large transformations, but changes according to
\begin{equation}
S_{CS}\rightarrow S_{CS}+2\pi kN,\ \ \ \ N \in \mathbb{Z}.
\label{action trans}
\end{equation}

The quantum theory can be defined by the partition function
\begin{equation}
Z=\int{\mathcal{D}A_{\mu}e^{\mathrm{i}S_{CS}}}.
\label{part function}
\end{equation}
Then, for a consistent quantization of the theory it is not $S$ that should be gauge invariant, but $\exp{\mathrm{i}S}$. From (\ref{action trans}), we obtain the quantization condition of the Chern-Simons level $k$,
\begin{equation}
k\in Z,
\end{equation}
in order to make the quantum theory gauge invariant.

Let us consider now the non-Abelian Chern-Simons theory defined on a manifold with a boundary. The presence of the boundary breaks even the small gauge transformations, which opens the possibility for describing dynamical modes at the edge. In fact, as discussed above, in the presence of a boundary, the action (\ref{non ab cs action}) is only invariant under a constrained gauge transformation with a group element satisfying $U(x)\xrightarrow{\partial \Omega} 1$. It is this lack of gauge invariance at the boundary that allows the emergence of dynamical edge modes. Whether there will be in fact dynamical modes on the boundary depends on the chosen boundary conditions for the fields. 

\begin{figure}[!h]
\centering
\includegraphics[scale=0.6]{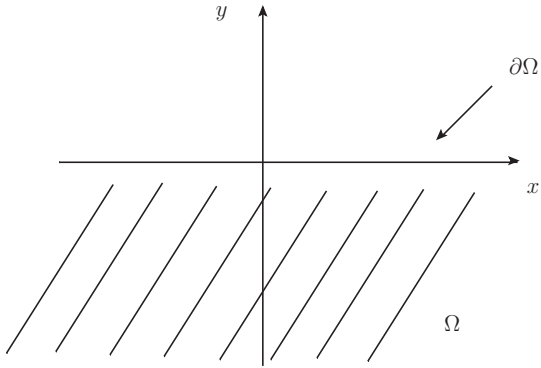}
\caption{Manifold with a boundary at $y=0$.}
\label{Boundary}
\end{figure}

To be concrete, we consider the system in the lower half $xy$ plane with boundary at $y=0$, as illustrate in Fig. \ref{Boundary}. The boundary conditions are chosen so that boundary terms do not contribute to the equations of motion. However, as one can verify, we still have some freedom and some choices indeed lead to dynamical edge modes. Particularly, the boundary condition
\begin{equation}
\left(A_t-vA_x\right)\Big|_{y=0}=0\label{BC}
\end{equation}
provides a dynamical chiral boson propagating at the edge. The parameter $v$ can be identified as the propagation velocity of the edge modes.
The condition (\ref{BC}) also guarantees that there are no contributions to the equations of motion from the boundary terms. It is worth mentioning that this condition is not a gauge fixing since the theory does not have gauge invariance at the boundary. Considering the constraint (\ref{BC}) in (\ref{non ab cs action}) we get
\begin{equation}
S_{CS}=\frac{k}{4\pi}\int_{\Omega}{d^3x\epsilon^{ij}\text{Tr}\left(A_tF_{ij}-A_i\partial_tA_j\right)}+\frac{k}{4\pi}\int_{\partial\Omega}{dtdx\text{Tr}v\left(A_x\right)^2},
\label{non ab cs action comp}
\end{equation}
where
\begin{equation}
F_{ij}=\partial_iA_j-\partial_jA_i+\left[A_i,A_j\right],\ \ i=x,y.
\label{field str}
\end{equation}

The quantum theory is defined by using this action in the partition function (\ref{part function}). Since there is no derivative acting on the component $A_t$ in (\ref{non ab cs action comp}), this is an auxiliary field and can be integrated out in the partition function. Performing this integral in $A_t$ we get a delta functional $\delta[F_{ij}]$, with $F_{ij}$ defined in eq. (\ref{field str}). Then, we have
\begin{equation}
Z=\int{\mathcal{D}A_{i}\delta[F_{ij}]\exp\left(\mathrm{i}\frac{k}{4\pi}\int_{\Omega}{d^3x\epsilon^{ij}\text{Tr}\left(-A_i\partial_tA_j\right)}+\mathrm{i}\frac{k}{4\pi}\int_{\partial\Omega}{dtdx\text{Tr}v\left(A_x\right)^2}\right)}.
\label{part function2}
\end{equation}

Let us introduce complex combinations of the coordinates in the $xy$ plane,
\begin{eqnarray}
&&z\equiv x+\mathrm{i}y~~~\text{and}~~~\bar{z}\equiv x-\mathrm{i}y\\
&&A_z\equiv \frac{1}{2}\left(A_x-\mathrm{i}A_y\right)~~~\text{and}~~~ A_{\bar{z}}\equiv \frac{1}{2}\left(A_x+\mathrm{i}A_y\right).
\label{complex coord}
\end{eqnarray}
In terms of these coordinates the partition function reads
\begin{equation}
Z=\int{\mathcal{D}A_{z}\mathcal{D}A_{\bar{z}}\delta[F_{z\bar{z}}]\exp\left(\frac{k}{4\pi}\int_{\Omega}{d^3x4\text{Tr}\left(-A_z\partial_tA_{\bar{z}}\right)}+\mathrm{i}\frac{k}{4\pi}\int_{\partial\Omega}{dtdx\text{Tr}v\left(A_z+A_{\bar{z}}\right)^2}\right)}.
\label{part function3}
\end{equation}
The solution of the constraint is a pure gauge field $A_i=-\partial_iMM^{-1}$, with $M\in SU(N)$. We can implement this constraint in the partition function above by first changing the variables according to the parametrization
\begin{eqnarray}
A_z&=&-\partial_zUU^{-1},\\
A_{\bar{z}}&=&U^{\dagger -1}\partial_{\bar{z}}U^{\dagger},
\label{fieds param}
\end{eqnarray}
where $U$ is a complex matrix with unit determinant. In the Appendix \ref{A2} we show that this change of variables introduces a Jacobian determinant that is canceled by another determinant coming from the same change of variables in the delta functional. Using these results we get
\begin{eqnarray}
Z&=&\int{\mathcal{D}U\mathcal{D}U^{\dagger}\delta[U^{-1}-U^{\dagger}]\exp\left(\frac{k}{4\pi}\int_{\Omega}{d^3x4\text{Tr}\left(\partial_zUU^{-1}\partial_t\left(U^{\dagger -1}\partial_{\bar{z}}U^{\dagger}\right)\right)}\right.}\\
&+&\mathrm{i}\left.\frac{k}{4\pi}\int_{\partial\Omega}{dtdx\text{Tr}v\left(-\partial_zUU^{-1}+U^{\dagger -1}\partial_{\bar{z}}U^{\dagger}\right)^2}\right).
\label{part function4}
\end{eqnarray}
By integrating on $U^{\dagger}$, changing back to real coordinates and after some algebra we obtain
\begin{eqnarray}
Z&=&\int{\mathcal{D}U\exp\left(\mathrm{i}\frac{k}{4\pi}\int_{\partial\Omega}{dxdt\text{Tr}\left(-\partial_xUU^{-1}\left(\partial_t-v\partial_x\right)UU^{-1}\right)}\right.}\nonumber\\
&+&\mathrm{i}\left.\frac{k}{12\pi}\int_{\Omega}{d^3x\epsilon^{\mu\nu\rho}\text{Tr}\left(\partial_{\mu}UU^{-1}U\partial_\nu U^{-1}U\partial_\rho U^{-1}\right)}\right).\label{part function5}
\end{eqnarray}
This partition function is the chiral WZW model already found in (\ref{c1.6}).  As is well known, the WZW model describes a class of 
two dimensional CFT's with rational central charges $c\geq 1$. 
As discussed previously, the WZW model is only a well defined quantum theory for integer $k$.

%%%%%%%%%%%%%%%%%%%%%%%%%%%%%%%%%%%%%%%%%%%%%%%
\subsection{Coset theories}

As emphasized above, the WZW model provides a field theory realization of the class of the $2D$ Rational CFT's with $c\geq 1$. In the WZW model, the Virasoro algebra, which is related to the conformal invariance of the theory, is derived from the  affine Lie algebra $\hat{g}_k$. We denote by $g$ the Lie algebra associated to the group $G$ of invariance of the WZW action and by $\hat{g}_k$ its affine extension of level $k$. From the algebraic perspective this structure is known as the Sugawara construction.

The Sugawara construction can be generalized to the so called Goddard-Kent-Olive (GKO) construction \cite{GKO}, where one considers a conformal theory described in terms of the Virasoro generators for a quotient of WZW models. Specifically, we consider an affine Lie algebra $\hat{g}$ and a $\hat{p}$ subalgebra of $\hat{g}$ and let $L^g_m$ and $L^p_m$ be the associated Virasoro generators. The quotient Virasoro generators are defined by $L^{g/p}_m\equiv L^g_m-L^p_m$ and they are shown to satisfy a Virasoro algebra by themselves. This coset construction is expected to complete the classification of all RCFT \cite{Seiberg}. The realization of the coset construction in terms of an action is given in terms of the gauged WZW model. 
%whereas the corresponding bulk Chern-Simons theory is described in \cite{Seiberg}.

We will review here the CS-WZW correspondence for the coset construction \cite{Fradkin}. We start by considering a simply connected compact Lie group $G$ and a subgroup $H$. Let $g$ and $h$ be, respectively, the associated Lie algebras and $A_{\mu}^aG^a$ and $B_\mu^iH^i$ are gauge field that take values in these algebras. We consider the following action
\begin{equation}
S[A,B]=k_gS_{CS}\left[A,G\right]-k_hS_{CS}\left[B,H\right].\label{coset action}
\end{equation}
The pair of non-Abelian CS actions $S_{CS}$
for the fields $A$ and $B$ are of the form (\ref{non ab cs action})
except for the CS levels $k_g$ and $k_h$, respectively, that have been factored out.
This pair of levels satisfy the relation
\begin{equation}
k_g = l k_h.
\end{equation}
The integer $l$ is called the embedding index of $h$ into $g$.

This action changes by boundary terms under the independent gauge transformations
\begin{eqnarray}
A^{\prime}_\mu&=&\Lambda^{-1}A_\mu \Lambda+ \Lambda^{-1}\partial_\mu \Lambda,\label{gauge trans A coset}\\
B^{\prime}_\mu&=&\omega^{-1}B_\mu \omega+ \omega^{-1}\partial_\mu \omega,\label{gauge trans B coset}
\end{eqnarray}
with $\Lambda$ and $\omega$ being elements of the gauge groups $G$ and $H$, respectively.

The dynamics on the boundary is obtained by fixing the following boundary conditions \cite{Fradkin}:
\begin{eqnarray}
\mathcal{P}_{h^\perp}\left(A_t+vA_x\right)&=&0,\label{bc perp h}\\
k_g\mathcal{P}_{g}\left(A_{\alpha}\right)-k_hB_{\alpha}&=&0, \ \ {\alpha}=t,x,\label{bc h}
\end{eqnarray}
where $\mathcal{P}_{h}$ and $\mathcal{P}_{h^\perp}$ are projectors on the subalgebra $h$ and on its orthogonal complement, respectively. The BC (\ref{bc perp h}) are analogous to the BC (\ref{BC}) and are chosen in order to provide a nontrivial dynamics at the boundary and to avoid edge contributions for the equations of motion for the vector fields, whereas the second set relates the components of the gauge fields that live in the same algebra subspace. It is this intertwining of the fields that provides a richer symmetry structure compared to the trivial extension $G\times H$.

By using (\ref{bc perp h}) and (\ref{bc h}) in the action (\ref{coset action}), we obtain
\begin{eqnarray}
Z&=&\int{\mathcal{D}A_\mu\mathcal{D}B_\nu\mathcal{D}\lambda\exp\left[\mathrm{i}\frac{k_g}{4\pi}\left(\int_{\Omega}{\epsilon^{ij}\text{Tr}_g\left(A_tF_{ij}-A_i\partial_tA_j\right)}+v\int_{\partial\Omega}{\text{Tr}_g\left(A^2_x\right)}\right)\right.}\nonumber\\
&-&\mathrm{i}\frac{k_h}{4\pi}\left.\left(\int_{\Omega}{\epsilon^{ij}\text{Tr}_h\left(B_tB_{ij}-B_i\partial_tB_j\right)}+v\int_{\partial\Omega}{\text{Tr}_h\left(B^2_x\right)}\right)\right.\nonumber\\
&+&\left.\mathrm{i}\frac{1}{2\pi}\int_{\partial\Omega}{\text{Tr}_h\left(\lambda\left(k_gA_x-k_hB_x\right)\right)}\right].\label{c1}
\end{eqnarray}
To write the partition function in this form we have used the fact that $\sum_{h^\perp}A^{a}_xA^a_x=\text{Tr}_g\left(A^2_x\right)-\text{Tr}_h\left(A^2_x\right)$ and the second set of BC (\ref{bc h}) to make $\text{Tr}_h\left(A^2_x\right)=\text{Tr}_h\left(B^2_x\right)$. Furthermore, since we still have the full $A_x$ field at the boundary, we inserted a Lagrange multiplier $\lambda^iH^i$, which is valued in the Lie-subalgebra $h$, to enforce the $x$ BC in (\ref{bc h}).

Now we can proceed in direct analogy with the non-Abelian case.
We integrate the non-dynamical fields
$A_t$ and $B_t$, which effectively play the role of auxiliary fields.
There follows the constraints
\begin{equation}
F_{ij}=0,\qquad B_{ij}=0,
\end{equation}
which are implemented by the pair of delta functions $\delta[F_{ij}]$ and $\delta[B_{ij}]$ that multiply the measure over $A_{i}$ and $B_{i}$, with $i=x,y$. As we discussed in the previous section, we can use complex parametrizations for the $A_i$and $B_i$ fields like in (\ref{complex coord}) with independent complex matrices $M$ and $M^{\dagger}$, and $N$ and $N^{\dagger}$, respectively. As we showed in the Appendix \ref{A2}, we can effectively replace the constrained measures $\mathcal{D}A_\mu\delta[F_{ij}]$ and $\mathcal{D}B_i\delta[B_{ij}]$ by $\mathcal{D}M$ and $\mathcal{D}N$, respectively, and use the solutions to the constraints, $A_i=-\partial_iMM^{-1}$ and $B_i=-\partial_iNN^{-1}$, into the action. We finally obtain
\begin{eqnarray}
Z&=&\int{\mathcal{D}M\mathcal{D}N\mathcal{D}\lambda\exp\left(\mathrm{i}k_gS^{Ch}_{WZW}[M]-\mathrm{i}k_hS^{Ch}_{WZW}[N]\right.}\nonumber\\
&+&\left.\mathrm{i}\frac{1}{2\pi}\int_{\partial\Omega}{\text{Tr}_h\left(\lambda\left(k_g\partial_xMM^{-1}-k_h\partial_xNN^{-1}\right)\right)}\right),
\end{eqnarray}
where $S^{Ch}_{WZW}$ is the chiral Wess-Zumino-Witten action defined in (\ref{part function5}).

For our purposes, we will consider the case of regular embeddings, that is $l=1$.
We now rewrite this partition function in a more convenient way that will allow us to make a more direct link to the quantum wire picture of the topological phases of matter. For this aim, let us perform the following change of variables \cite{Fradkin}:
\begin{equation}
U=MN^{-1},\ \ C_x=-\partial_xNN^{-1},\ \ C_t=\lambda\label{change var}
\end{equation}
and use the Polyakov-Wiegmann identity \cite{Nair} for the chiral action,
\begin{equation}
S^{ch}_{WZW}[MN]=S^{ch}_{WZW}[M]+S^{ch}_{WZW}[N]-\frac{1}{2\pi}\int{d^2x\text{Tr}\left(M^{-1}\partial_{-}M\partial_{x}NN^{-1}\right)}.\label{PW identity}
\end{equation}
Using these results and factoring the group volume $\mathcal{D}h$, we obtain
\begin{eqnarray}
Z&=&\int{\mathcal{D}U\mathcal{D}C_i\exp\left(\mathrm{i}k_gS^{Ch}_{WZW}[U]\right.}\nonumber\\
&+&\left.{\mathrm{i}\frac{k_g}{2\pi}\int_{\partial\Omega}{\text{Tr}_h\left(-C_xU^{-1}\partial_-U+C_t\partial_xUU^{-1}-C_tUC_xU^{-1}+C_tC_x\right)}}\right).\label{c2}
\end{eqnarray}
In the change of variable from $N$ to $C_x$ we should account a non-trivial Jacobian $\det(D_x)$. However, as pointed in \cite{Seiberg}, by a canonical treatment one can show that this Jacobian does not contribute. We recognize the action in this expression as the gauged version of the chiral WZW model. In this case, we start with the chiral WZW model for a group valued field $U\in G$ and gauge a subgroup $H$ of $G$ by coupling a dynamical gauge field $C_i\in h$ to $U$ in an anomaly free way.

%%%%%%%%%%%%%%%%%%%%%%%%%%%%%%%%%%%%%%%%%%%%%%%%%%%%%%%%%%%%%

\subsection{Coset model for $\left(\frac{U(2k)}{U(1)\times SU(k)_2}\times \frac{U(2k')}{U(1)\times SU(k')_2}\right)/{SU(2)_{k+k'}}$}

In this section we will use the results of our previous analysis to construct the connection between the bulk Chern-Simons theory and the boundary WZW model for the specific coset of our interest in this paper. We then will be able to associate an effective Chern-Simons theory to the description of the topological phases in terms of the quantum wires.

We will denote the fields in the direct sums of algebras with a tilde, whereas the pairs of fields in the two separate spaces will be differed by a prime. To start with we need to consider a gauge field $\tilde{\Omega}_\mu$ that takes values in the direct sum of Lie algebras $u(2k)\bigoplus u(2k')$. Then $\tilde{\Omega}_\mu$ can also be written as a direct sum of gauge fields $\Omega_\mu\bigoplus\Omega'_\mu$, with $\Omega_\mu\in u(2k)$ and $\Omega_\mu\in u(2k')$. It is also convenient to extract the $u(1)$ gauge fields $a_\mu$ and $a'_\mu$ from the gauge fields $\Omega_\mu$ and $\Omega'_\mu$, respectively. With $\tilde{a}_\mu=a_\mu\bigoplus a'_\mu$ we can write the Kronecker sum $\tilde{\Omega}_\mu=\tilde{a}_\mu\oplus\tilde{\beta}_\mu$, with $\tilde{\beta}\in su(2k)\bigoplus su(2k')$. The corresponding $su(2k)$ and $su(2k')$-fields are denoted by $\beta_\mu$ and $\beta'_\mu$, respectively. Explicitly, we have
\begin{equation}
\tilde{\Omega}_\mu=\Omega_\mu\bigoplus\Omega'_\mu=a_\mu\oplus\beta_\mu\bigoplus a'_\mu\oplus\beta'_\mu.\label{d0}
\end{equation}

To get the aimed coset structure we introduce a gauge field $\tilde{\mathcal{A}}_\mu$ belonging to the subalgebra $\left(u(1)\oplus su(k)_2\bigoplus u(1)\oplus su(k')_2\right)\oplus su(2)_{k+k'}$. In terms of fields in the smaller subalgebras, we have
\begin{equation}
\tilde{\mathcal{A}}_\mu=A_\mu\oplus B_\mu\oplus C_\mu\bigoplus A'_\mu\oplus B'_\mu\oplus C_\mu,\label{d1}
\end{equation}
with $B_\mu$ and $B'_\mu$ $\in$ $su(k)$ and $su(k')$, respectively, and $C_\mu\in su(2)$.

According to the general discussion of the previous section, the CS theory for this coset is then given by
\begin{equation}
S=S_{CS}\left[\tilde{\Omega},G\right]-S_{CS}\left[\tilde{\mathcal{A}},H\right],\label{d1.1}
\end{equation}
suplemented by the boundary conditions
\begin{eqnarray}
\mathcal{P}_{h^\perp}\left(\tilde{\Omega}_t+v\tilde{\Omega}_x\right)&=&0,\label{d2}\\
\mathcal{P}_{h}\left(\tilde{\Omega}_\alpha\right)-\tilde{\mathcal{A}}_\alpha&=&0, \ \ \alpha=t,x.\label{d3}
\end{eqnarray}
Following the same steps from (\ref{c1}) to (\ref{c2}) we obtain the boundary chiral WZW partition function
\begin{eqnarray}
Z&=&\int{\mathcal{D}\tilde{w}\mathcal{D}\tilde{\mathcal{A}}_i\exp\left(\mathrm{i}S^{Ch}_{gauge}[\tilde{w},\tilde{\mathcal{A}}]\right)}\nonumber\\
&=&\int{\mathcal{D}\tilde{w}\mathcal{D}\tilde{\mathcal{A}}_i\exp\left(\mathrm{i}S^{Ch}_{WZW}[\tilde{w}]\right.}\nonumber\\
&+&\left.{\frac{\mathrm{i}}{2\pi}\int_{\partial\Omega}{\text{Tr}_H\left(-\tilde{\mathcal{A}}_x\tilde{w}^{-1}\partial_-\tilde{w}+\tilde{\mathcal{A}}_t\partial_x\tilde{w}\tilde{w}^{-1}-\tilde{\mathcal{A}}_t\tilde{w}\tilde{\mathcal{A}}_x\tilde{w}^{-1}+\tilde{\mathcal{A}}_t\tilde{\mathcal{A}}_x\right)}}\right),\label{d4}
\end{eqnarray}
where $\tilde{w}\in u(2k)\times u(2k')$ and can be further decomposed as
\begin{equation}
\tilde{w}=w\bigoplus w'=\chi\otimes v\bigoplus\chi'\otimes v',\label{d5}
\end{equation}
with $\chi$ and $\chi'$ $\in$ $u(1)$ and $v$ ($v'$) $\in$ $su(2k)$ ($su(2k')$).
The gauge fields are Lagrange multipliers that ensure the constraints of vanishing subgroup currents. In terms of the fields in the smaller subalgebras we can write the partition function (\ref{d4}) as
\begin{eqnarray}
Z&=&\int{\mathcal{D}v\mathcal{D}v'\mathcal{D}B_i\mathcal{D}C_i\exp\left\{\mathrm{i}\left[S^{ch}[v]+S^{ch}[v']\right.\right.}\nonumber\\
&+&\left.\left.\frac{1}{\pi}\int_{\Omega}{\text{Tr}\left(B_t\partial_xvv^{-1}+B_xv^{-1}\partial_-v-B_tvB_xv^{-1}-B_tvC_xv^{-1}+B_tB_x\right)}\right.\right.\nonumber\\
&+&\left.\left.\frac{1}{\pi}\int_{\Omega}{\text{Tr}\left(B'_t\partial_xv'v'^{-1}+B'_xv'^{-1}\partial_-v'-B'_tv'B'_xv'^{-1}-B'_tv'C_xv'^{-1}+B'_tB'_x\right)}\right.\right.\nonumber\\
&+&\frac{1}{2\pi}\left.\left.\int_{\Omega}{\text{Tr}\left(k\left(C_t\partial_xvv^{-1}+C_xv^{-1}\partial_-v-C_tvC_xv^{-1}-C_tvB_xv^{-1}+C_tC_x\right)\right.}\right.\right.\nonumber\\
&+&\left.\left.\left.k'\left(C_t\partial_xv'v'^{-1}+C_xv'^{-1}\partial_-v'-C_tv'B'_xv'^{-1}-v'C_xv'^{-1}+C_tC_x\right)\right)\right]\right\}.\label{d7}
\end{eqnarray}
We can notice that the $U(1)$ fields $\chi$ and $\chi'$ were withdrawn from the partition function due to the constraints imposed by the abelian gauge fields $A$ and $A'$, which are likewise suppressed.

It is also convenient to express the bulk theory (\ref{d1.1}) in terms of the fields in the smaller subalgebras that compose the whole coset structure. From (\ref{d0}), (\ref{d1}), and (\ref{d1.1}) we obtain
\begin{eqnarray}
S&=&\frac{1}{4\pi}\int{\epsilon^{\mu\nu\rho}\left(a_\mu\partial_\nu a_\rho\right)}+\frac{1}{4\pi}\int{\epsilon^{\mu\nu\rho}\left(a'_\mu\partial_\nu a'_\rho\right)}\nonumber\\
&+&\frac{1}{4\pi}\int{\epsilon^{\mu\nu\rho}\text{Tr}_{su(2k)}\left(\beta_\mu\partial_\nu\beta_\rho+\frac{2}{3}\beta_\mu\beta_\nu\beta_\rho\right)}+\frac{1}{4\pi}\int{\epsilon^{\mu\nu\rho}\text{Tr}_{su(2k')}\left(\beta'_\mu\partial_\nu\beta'_\rho+\frac{2}{3}\beta'_\mu\beta'_\nu\beta'_\rho\right)}\nonumber\\
&-&\frac{1}{4\pi}\int{\epsilon^{\mu\nu\rho}\left(A_\mu\partial_\nu A_\rho\right)}-\frac{1}{4\pi}\int{\epsilon^{\mu\nu\rho}\left(A'_\mu\partial_\nu A'_\rho\right)}\nonumber\\
&-&\frac{1}{2\pi}\int{\epsilon^{\mu\nu\rho}\text{Tr}_{su(k)}\left(B_\mu\partial_\nu B_\rho+\frac{2}{3}B_\mu B_\nu B_\rho\right)}-\frac{1}{2\pi}\int{\epsilon^{\mu\nu\rho}\text{Tr}_{su(k')}\left(B'_\mu\partial_\nu B'_\rho+\frac{2}{3}B'_\mu B'_\nu B'_\rho\right)}\nonumber\\
&-&\frac{k+k'}{4\pi}\int{\epsilon^{\mu\nu\rho}\text{Tr}_{su(2)}\left(C_\mu\partial_\nu C_\rho+\frac{2}{3}C_\mu C_\nu C_\rho\right)},\label{d8}
\end{eqnarray}
which we need to supplement with the following boundary conditions:
\begin{eqnarray}
a_\alpha&=&A_\alpha,\label{d9}\\
\mathcal{P}^{\perp}_{su(k)\oplus su(2)}\left(\beta_t+v\beta_x\right)&=&0,\label{d10}\\
\mathcal{P}^{\perp}_{su(k')\oplus su(2)}\left(\beta'_t+v\beta'_x\right)&=&0,\label{d11}\\
\mathcal{P}_{su(k)}\left(\beta_\alpha\right)&=&B_\alpha,\label{d12}\\
\mathcal{P}_{su(k')}\left(\beta'_\alpha\right)&=&B'_\alpha,\label{d13}\\
{P}_{su(2)}\left(\beta_\alpha+\beta'_\alpha\right)&=&C_\alpha,\label{d14}
\end{eqnarray}
with $\alpha=t,x$.

In Sec. \ref{CGWZ} we have showed that the edge theory for the low-energy dynamical modes coincides with the gauged chiral WZW model. Similarly, we showed in this section that the bulk low-energy effective field theory corresponds to the same gauged WZW model. Therefore, we can conclude that the Chern-Simons theory (\ref{d8}) supplemented with the boundary conditions (\ref{d9})-(\ref{d14}) can be associated to the low-energy effective theory for the interacting electrons propagating in the quantum wires.

%%%%%%%%%%%%%%%%%%%%%%%%%%%%%%%%%%%%%%%%%%%%%%%%%%

%%%%%%%%%%%%%%%%%%%%%%%%%%%%%%%%%%%%%%%%%%%%%%%%%%
\section{Summary}\label{Sec6}

In spite of the achievements toward a complete characterization of two dimensional rational conformal field theories in terms of 2+1 dimensional topological Chern-Simons theories, a general classification of non-Abelian topological ordered phases of matter in terms of topological effective field theories is still missing. In this work we discussed how to implement this connection for a class of non-Abelian phases, namely the class of non-Abelian topological spin liquids characterized by chiral edge states with central charge corresponding to the coset structure $su(2)_k\oplus su(2)_{k'}/su(2)_{k+k'}$. This class was recently predicted in the work of Ref. \cite{Huang} by analyzing the strong coupling limit of a system of interacting quantum wires. 

In our pursuance of the effective low-energy field theory for the coset structure, we also argued in favor of the conjectured stability of the strongly coupled fixed point. Firstly, we started with the interacting quantum wires system and by taking its strong coupling limit we obtained the partition function of the constrained free fermions  after integrating over decoupled gauge fields.
In this way, we showed that the system of constrained free fermions (\ref{2.253a}) may realize the conjectured strongly coupled topological phase of the interacting quantum wires by matching the central charges of both systems. The gapless sector of the constrained fermion system, located at the edges, was then shown to be equivalent to a chiral gauged WZW models at the boundaries, which in turn was connected with a bulk Chern-Simons theory through the bulk-edge correspondence. In Sec. \ref{Sec3} we explicitly constructed the bulk Chern-Simons theory as well as the boundary conditions for the bulk gauge fields to get the desired central charge for the boundary coset conformal field theory. The effective theory can be used in principle to explore interesting properties
of the topological phase, like the braiding statistics of the non-Abelian anyons and the topological degeneracy of the ground state. We also expect that the construction done in this work can be extended to other classes of topological phases with important phenomenological consequences.

%%%%%%%%%%%%%%%%%%%%%%%%%%%%%%%%%%%%%%%%%%%%%%%%%%

\section{Acknowledgments}

We would like to thank Claudio Chamon and Christopher Mudry for the careful reading of the manuscript, the constructive criticism and 
for sharing with us their manuscript prior publication. 
We acknowledge the financial support of Brazilian agency CAPES.

%%%%%%%%%%%%%%%%%%%%%%%%%%%%%%%%%%%%%%%%%%%%%%%%%%%%
\appendix

\section{Integration over Fermions}\label{A1}

We want to show the result (\ref{3.25}), which we reproduce here for convenience
\begin{eqnarray}
&&\int \mathcal{D}\psi_R \mathcal{D}\psi_L \exp \mathrm{i} \int dt dx \,\mathcal{\tilde{L}}[\psi_R,\psi_L, A_+,A'_+,B_+,B'_+,C_+]\nonumber\\
&=& \det \partial_{+}\det \partial_{-} \exp  \mathrm{i}\left(- S[\phi] - S[\phi'] -  2 S[h] - 2S[h'] - (k+k') S[g] \right).
\label{a1}
\end{eqnarray}
Let us start with the Langrangian (\ref{3.10}) with the gauge fixing conditions (\ref{3.18}), (\ref{3.19}) and (\ref{3.20}). 
Remember the parametrization for the remaining gauge fields (\ref{3.22}), (\ref{3.23}) and (\ref{3.24}). 
Now define the fields $U$ and $U'$ which are the direct product of 
\begin{equation}
U\equiv \phi\otimes h\otimes g~~~\text{and}~~~U'\equiv \phi'\otimes h'\otimes g.
\label{a2}
\end{equation}
Note that the $g$-field is the same in both expressions, since it is the parametrization of the diagonal field. 
Thus it follows that
\begin{equation}
U^{-1}\partial_+U=\phi^{-1}\partial_{+}\phi \oplus h^{-1}\partial_{+}h\oplus g^{-1}\partial_{+}g
\label{a3}
\end{equation}
and
\begin{equation}
U^{\prime -1}\partial_+U'=\phi^{\prime -1}\partial_{+}\phi' \oplus h^{\prime -1}\partial_{+}h' \oplus g^{-1}\partial_{+}g.
\label{a4}
\end{equation}
With this the first line in (\ref{a1}) can be integrated out to produce
\begin{eqnarray}
&&\int \mathcal{D}\psi_R \mathcal{D}\psi_L \exp \mathrm{i} \int dt dx\,\mathcal{\tilde{L}}[\psi_R,\psi_L, A_+,A'_+,B_+,B'_+,C_+]\nonumber\\
&=&\det(\partial_+ + \mathrm{i} U^{-1}\partial_+U)\det\partial_-\det(\partial_+ + \mathrm{i} U^{\prime -1}\partial_+U')\det\partial_-\nonumber\\
&=&\underbrace{\det\partial_+\det\partial_-}_{k~ \text{times}}\underbrace{\det\partial_+\det\partial_-}_{k'~\text{times}}\exp \mathrm{i} (-S[U]-S[U']),
\label{a5}
\end{eqnarray}
where $S$ is the WZW action given in (\ref{3.26}). 
We can then use the Polyakov-Wiegmann identity (\ref{3.44}) to write $S[U]$ and $S[U']$ in terms of original fields of (\ref{a2}). So it is easy to get
\begin{equation}
S[U]= 2k S[\phi]+2S[h]+kS[g]~~~\text{and}~~~S[U']= 2k' S[\phi']+2S[h']+kS[g],
\label{a7}
\end{equation}
where we have used that $S[f\otimes I]=\text{Tr}I S[f]$ and that $\text{Tr} f^{-1}\partial f=0$, 
for any $f\in SU(N)$, since $f^{-1}\partial f\in su(N)$ and the trace of generators vanishes.  
Note that the factors multiplying the Abelian actions $S[\phi]$ and $S[\phi']$ can be absorbed into a 
redefinition of the fields. For example, they disappear if we write $\phi=e^{\frac{\mathrm{i}}{\sqrt{2k}}\varphi}$ and  $\phi'=e^{\frac{\mathrm{i}}{\sqrt{2k'}}\varphi'}$.
This is not true for the non-Abelian contributions.   
Plugging these results in (\ref{a5}) yields to (\ref{a1}).

%%%%%%%%%%%%%%%%%%%%%%%%%%%%%%%%%%%%%%%%%
\section{Constrained Measure}\label{A2}

Let us consider the constrained measure
\begin{equation}
\mathcal{D}A_i\delta[F_{ij}].\label{constr measure}
\end{equation}
In terms of the complex variables
\begin{eqnarray}
&&z=x+\mathrm{i}y,\ \ \bar{z}=x-\mathrm{i}y\\
&&A_z=\frac{1}{2}\left(A_x-\mathrm{i}A_y\right),\ \ A_{\bar{z}}=\frac{1}{2}\left(A_x+\mathrm{i}A_y\right),\label{complex coord1}
\end{eqnarray}
the measure (\ref{constr measure}) is written as
\begin{equation}
\mathcal{D}A_z\mathcal{D}A_{\bar{z}}\delta[F_{z\bar{z}}].\label{constr measure complex}
\end{equation}
Now, we perform the change of variables
\begin{eqnarray}
&&A_z=-\partial_zUU^{-1},\\
&&A_{\bar{z}}=U^{\dagger -1}\partial_{\bar{z}}U^{\dagger}.\label{repar}
\end{eqnarray}
Since,
\begin{eqnarray}
&&\delta A_z=-D_z\left(\delta UU^{-1}\right),\\
&&\delta A_{\bar{z}}=D_{\bar{z}}\left(U^{\dagger -1}\delta U^{\dagger}\right),\label{var}
\end{eqnarray}
the change of variables (\ref{repar}) changes the measure $\mathcal{D}A_z\mathcal{D}A_{\bar{z}}$ as it follows:
\begin{equation}
\mathcal{D}A_z\mathcal{D}A_{\bar{z}}=\left|\det\left(D_zD_{\bar{z}}\right)\right|\mathcal{D}U\mathcal{D}U^{\dagger}.\label{jacobian}
\end{equation} 

The delta functional can also be rewritten in terms of the new variables $U$ and $U^{\dagger}$ by noticing first that
\begin{eqnarray}
\delta F_{z\bar{z}}&=&-D_{\bar{z}}\delta A_z+D_z\delta A_{\bar{z}},\nonumber\\
&=&D_{\bar{z}}D_z\left(\delta UU^{-1}\right)+D_zD_{\bar{z}}\left(U^{\dagger -1}\delta U^{\dagger}\right).\label{var fs}
\end{eqnarray}
By using the property of the delta function
\begin{equation}
\delta(f(x,y))=\frac{\delta(x-x_0(y)))}{\left|\partial_xf(x_0(y),y)\right|},\label{delta prop}
\end{equation}
we have
\begin{equation}
\delta[F_{z\bar{z}}]=\frac{\delta[U^{-1}-U^{\dagger}]}{\left|\det\left(D_{\bar{z}}D_z\right)\right|},\label{delta prop funct}
\end{equation}
since the solution to the constraint $F_{z\bar{z}}=0$ is $U^{-1}=U^{\dagger}$.
From (\ref{constr measure complex}), (\ref{jacobian}) and (\ref{delta prop funct}) we finally have
\begin{eqnarray}
\mathcal{D}A_z\mathcal{D}A_{\bar{z}}&=&\left|\det\left(D_zD_{\bar{z}}\right)\right|\frac{\delta[U^{-1}-U^{\dagger}]}{\left|\det\left(D_{\bar{z}}D_z\right)\right|}\mathcal{D}U\mathcal{D}U^{\dagger}\nonumber\\
&=&\delta[U^{-1}-U^{\dagger}]\mathcal{D}U\mathcal{D}U^{\dagger}.\label{measure new var}
\end{eqnarray}
It is this canceling of determinants that justify the change from the constrained measure in (\ref{part function3}) to the one that appears in (\ref{part function4}).

%%%%%%%%%%%%%%%%%%%%%%%%%%%%%%%%%%%%%%%%%%%%%%%%%%%%%%%

%%%%%%%%%%%%%%%%%%%%%%%%%%%%%%%%%%%%%%%%%%%%%%%%%%%%%%%
\end{document}